\documentclass[aps,twocolumn,floatfix,superscriptaddress]{revtex4}

\usepackage{graphicx}
\usepackage{amsfonts}
\usepackage{amssymb}
\usepackage{amsbsy}
\usepackage{amsmath}
\usepackage{mathrsfs}
\usepackage{latexsym}
\usepackage{natbib}
\usepackage{bm}
\usepackage{subfigure} 
\usepackage{color}
\usepackage{wasysym}
\usepackage{mathbbol}
\usepackage{bigints}
\usepackage{hyperref}
\allowdisplaybreaks
\usepackage{comment}

\definecolor{napiergreen}{rgb}{0.16, 0.5, 0.0}

\renewcommand*{\l}{\lambda_{\star}}

\def\nn{\nonumber} 
\def\pa{{\partial}}
\def\f{\frac}
\def\l{\left}
\def\r{\right}
\def\d{{\rm d}}
\def\tx{\tilde{x}}
\def\om{0_{_\mathrm{M}}}

\def\cE{\mathcal{E}}
\def\cEb{\bar{\mathcal{E}}}

\begin{document}

\title{Radiative processes of single and entangled detectors on\\
circular trajectories in $(2+1)$~dimensional Minkowski spacetime}
\author{Subhajit Barman}
\email{subhajit.barman@physics.iitm.ac.in}
\affiliation{Centre for Strings, Gravitation and Cosmology,
Department of Physics, Indian Institute of Technology Madras, 
Chennai 600036, India}
\affiliation{Department of Physics, Indian Institute of Technology Guwahati, 
Guwahati 781039, Assam, India}
\author{Bibhas Ranjan Majhi}
\email{bibhas.majhi@iitg.ac.in}
\affiliation{Department of Physics, Indian Institute of Technology Guwahati, 
Guwahati 781039, Assam, India}
\author{L. Sriramkumar}
\email{sriram@physics.iitm.ac.in}
\affiliation{Centre for Strings, Gravitation and Cosmology,
Department of Physics, Indian Institute of Technology Madras, 
Chennai 600036, India}
\begin{abstract}
We investigate the radiative processes involving two entangled Unruh-DeWitt 
detectors that are moving on circular trajectories in $(2+1)$-dimensional 
Minkowski spacetime.
We assume that the detectors are coupled to a massless, quantum scalar field,
and calculate the transition probability rates of the detectors in the Minkowski 
vacuum as well as in a thermal bath.
We also evaluate the transition probability rates of the detectors when they 
are switched on for a finite time interval with the aid of a Gaussian switching
function.
We begin by examining the response of a single detector before we go on to 
consider the case of two entangled detectors.
As we shall see, working in $(2+1)$ spacetime dimensions makes the computations 
of the transition probability rates of the detectors relatively simpler. 
We find that the cross transition probability rates of the two entangled detectors
can be comparable to the auto transition probability rates of the individual 
detectors. 
We discuss specific characteristics of the response of the entangled detectors 
for different values of the parameters involved and highlight the effects of the 
thermal bath as well as switching on the detector for a finite time interval.
\end{abstract}
\maketitle


\section{Introduction}\label{Introduction}

In quantum field theory in Minkowski spacetime, the concept of a particle 
is covariant under Lorentz transformations.
However, about half-a-century ago, it was discovered that the notion of a
particle is not a generally covariant concept (for the original discussion,
see Ref.~\cite{Fulling:1972md}; for detailed discussions, see the 
textbooks~\cite{Birrell:1982ix,Fulling:1989nb,Mukhanov:2007zz,Parker:2009uva}).
In general, an observer in motion along a non-inertial trajectory 
in flat spacetime may see the Minkowski vacuum to be populated 
with particles~\cite{Letaw:1980yv,Letaw:1980ik,Sriramkumar:1999nw}.
For instance, a uniformly accelerated observer sees the Minkowski vacuum 
as a thermal bath, a phenomenon that has come to be known as the Unruh 
effect (for the original discussion, see Refs.~\cite{Unruh:1976db,DeWitt:1980hx}; 
for a detailed review on the phenomenon, see, for instance, 
Ref.~\cite{Crispino:2007eb}).

Over the last few decades, there has been a constant effort to understand
the notion of a particle in a curved spacetime.
The idea of detectors was originally introduced to provide an operational
definition to the concept of a particle~\cite{Unruh:1976db,DeWitt:1980hx,
Letaw:1980ik}.
By a detector one has in mind, say, a two level system which interacts 
with the quantum field of interest and is excited or de-excited when it 
is in motion.
The response of detectors that are in motion on a variety of trajectories and 
are coupled to the quantum field in different manner have been examined in flat 
and curved spacetimes (for an inexhaustive list, see Refs.~\cite{Letaw:1980ik,
Bell:1982qr,Svaiter:1992xt,Higuchi:1993cya,Sriramkumar:1994pb,Davies:1996ks,
Sriramkumar:1999nw,Korsbakken:2004bv,Gutti:2010nv,JaffinoStargen:2017qyh,
Louko:2017emx,Costa:2020aqa,Biermann:2020bjh}.)

At this stage, we should clarify that, in general, the response of the
detectors may not match the results obtained from more formal methods
such as the Bogoliubov transformations and the effective Lagrangian,
which also reflects the particle content of the field (for a discussion
in this context, see Ref.~\cite{Sriramkumar:2016nmn}).
Moreover, apart from depending on the trajectory, the response of the 
detectors depends on the nature of their interaction with the quantum field. 
Nevertheless, the response of the detectors has been studied extensively
in a variety of situations. 
In particular, it has been recognized that the idea of detectors can prove 
to be indispensable to experimentally observe the phenomenon of the Unruh 
effect or its equivalents (in this context, see, for example,
Refs.~\cite{Schutzhold:2006gj,Sudhir:2021lpf}).
Therefore, it seems important to construct specific models of detectors 
which closely capture possible experimental realizations and investigate
the response of these detectors under different conditions.

In the literature, we find that a significant amount of attention has been 
paid to detectors that are in uniformly accelerated motion.
Evidently, this interest has been due to the fact that uniformly 
accelerated detectors exhibit a thermal response, which has a close 
analogy with Hawking radiation from black holes~\cite{Birrell:1982ix,
Mukhanov:2007zz}.
But, from a practical and experimental perspective, it seems more convenient 
to consider detectors that are moving on circular trajectories (for early 
discussions, see Refs.~\cite{Bell:1982qr,Davies:1996ks,Korsbakken:2004bv}; 
for more recent discussions in this context, see
Refs.~\cite{Gutti:2010nv,JaffinoStargen:2017qyh,Louko:2017emx,Biermann:2020bjh}).
Moreover, often the response of the detectors has been evaluated assuming
that they remain switched on for infinite time.
Needless to say, if such non-trivial phenomena are to be experimentally observed, 
it becomes important to examine the response of detectors that are switched
on for a finite time interval.

With the above motivations in mind, in this work, we examine the response
of the so-called Unruh-DeWitt detectors that are coupled to a massless,
quantum scalar field through a monopole interaction and are in motion on 
circular trajectories in Minkowski spacetime.
The Unruh-DeWitt monopole detectors are the simplest of the different possible
detectors in the sense that they are coupled linearly to the quantum 
field~\cite{Unruh:1976db,DeWitt:1980hx}.
We evaluate the infinite time as well as the finite time response of these
detectors.
We shall work with Gaussian window functions to switch the detectors on 
for a finite time interval (for early discussions in this context, see
Ref.~\cite{Svaiter:1992xt,Higuchi:1993cya,Sriramkumar:1994pb}; for recent 
discussions, see Ref.~\cite{Costa:2020aqa}).
For mathematical convenience, we shall work in $(2+1)$-spacetime dimensions, 
and calculate the transition probability rate of the detectors in the 
Minkowski vacuum and in a thermal bath. 
We should mention that our focus on the $(2+1)$-dimensional case is also 
motivated by its extensive consideration in models of analogue gravity (in 
this regard, see Ref.~\cite{Sanchez-Kuntz:2022gds} and the references therein).
After discussing the case of a single detector, we shall go on to calculate
the transition probability rate of two detectors that are assumed to be in 
an entangled initial state, a situation that has drawn considerable attention 
in the literature over the last few years (in this regard, see, for example,
Refs.~\cite{Hu:2015lda,Menezes:2015uaa,Arias:2015moa,Flores-Hidalgo:2015urj, 
Menezes:2015iva,Menezes:2015veo,Rizzuto:2016ijj,Rodriguez-Camargo:2016fbq, 
Zhou:2016urt,Cai:2017jan,Liu:2018zod,Zhou:2020oqa,Barman:2021oum}; for a 
discussion on entangled detectors in circular motion, see, for instance,
Refs.~\cite{Costa:2020aqa,Zhang:2020xvo}).
We shall focus on the excitation of the detector (i.e. it absorbs than emits 
quanta) due to its interaction with the quantum field and its motion.
As we shall illustrate, when the detectors are in circular motion and are 
switched on for an infinite time interval, generically, the transition 
probability rate of the detectors in the Minkowski vacuum and in the thermal 
bath is higher when the energy gap between the two levels of the detectors 
is smaller and the velocity of the detector is larger.
We also find that, in a thermal bath, when the detectors remain switched on 
for an infinite time interval, the higher the temperature of the bath, the
higher is response of the detectors.
Interestingly, we find that the transition probability rate of the detectors 
in the Minkowski vacuum are higher when they are switched on for a shorter
time interval, and we should point out that similar phenomenon has also been 
noticed previously in the literature (see, for instance, 
Refs.~\cite{Sriramkumar:1994pb,Louko:2006zv,Satz:2006kb,Brown:2012pw}). 
The corresponding transition probability rate in a thermal bath exhibits a 
more complex behavior, with the transition probability rate being higher when 
the temperature is higher provided the energy gap is large, while the behavior 
can be reversed for lower energy gaps, depending on the temperature. 
We also discuss different aspects of the total transition probability rate of 
the detectors for specific transitions from the symmetric and anti-symmetric 
entangled states to the collective excited state due to the presence of the 
thermal bath and the Gaussian switching function.

This paper is organized as follows.
In Sec.~\ref{sec:ed-gd}, we shall introduce and describe the response of 
two entangled Unruh-DeWitt detectors that are in motion along specific
trajectories and are interacting with a quantum scalar field.
In Sec.~\ref{sec:rrd-mv-tb}, we shall discuss the response of a {\it single}\/
Unruh-DeWitt detector that is moving on a circular trajectory in 
$(2+1)$-dimensional Minkowski spacetime.
We shall evaluate the transition probability rates of the detector in the
Minkowski vacuum as well as in a thermal bath.
We shall also consider the finite time transition probability rates of these 
detectors when they are switched on and off with the help of a Gaussian 
switching function.
As we shall see, these calculations for a single detector prove to be helpful 
later when we evaluate the responses of the entangled detectors.
In Sec.~\ref{sec:erd-mv-tb}, we shall evaluate the auto and cross transition
probability rates of two entangled detectors that are in motion along circular
trajectories, when the field is assumed to be in the Minkowski vacuum and in a
thermal bath.
We shall also discuss the response of these entangled detectors when they are
switched on for a finite time interval.
We shall conclude in Sec.~\ref{sec:sd} with a summary of the results we have 
obtained and a discussion on the broader implications of our analysis.
We shall relegate some of the additional discussions to the appendices.

A brief word on our notation is in order at this stage of our discussion.
We shall work with units such that $\hbar=c=1$.
For convenience, we shall describe the set of spacetime coordinates $(t,
{\bm x})$ collectively as~$\tx$. 


\section{Radiative processes of two entangled detectors:~The model}\label{sec:ed-gd}

In this section, we shall briefly outline the radiative processes that 
arise in situations involving two entangled Unruh-DeWitt detectors.
The discussion allows us to introduce the notation and also describe 
the quantities that we shall evaluate later.
We should mention that the model we shall consider has been examined 
earlier in different situations (in this context, see, for instance,
Refs.~\cite{Arias:2015moa,Rodriguez-Camargo:2016fbq,Costa:2020aqa}). 

\begin{figure}[!t]
\centering
\includegraphics[width=1\linewidth]{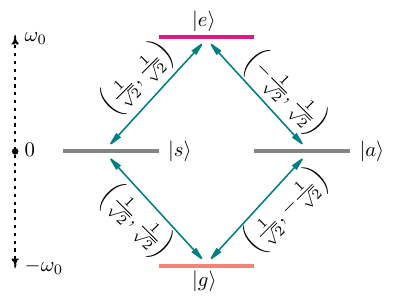}
\caption{An illustration of the energy levels corresponding to the
eigen states of the two entangled detectors with two levels each. 
(We should mention that this figure has been taken from Ref.~\cite{Costa:2020aqa}.)
The contributions from the monopole moment for each transition have
also been indicated in the figure.}\label{fig:Energy-levels}
\end{figure}

The detectors we shall consider are assumed to be composed of point like 
atoms with two internal energy levels, which are interacting with a scalar 
field through a monopole interaction.
For simplicity, we shall assume the field to be a massless, minimally 
coupled real scalar field, say,~$\Phi$. 
The Hamiltonian of the complete system composed of the two detectors and 
the scalar field is assumed to be of the form
\begin{equation}\label{eq:Hamiltonian-total}
H = H_\mathrm{D}+H_\mathrm{F}+H_\mathrm{I},
\end{equation}
where $H_\mathrm{D}$ denotes the Hamiltonian of the
detectors free of any interaction, $H_\mathrm{F}$ is the Hamiltonian 
describing the free scalar field, and the term $H_\mathrm{I}$ describes 
the interaction between the detectors and the scalar field. 
As initially suggested by Dicke (for the original discussion, see 
Ref.~\cite{Dicke:1954}; for a recent discussion, see 
Ref.~\cite{Rodriguez-Camargo:2016fbq}), one may express the Hamiltonian 
describing the two static atoms constituting the detectors as follows:
\begin{equation}
H_\mathrm{D} 
= \omega_{0}\, \l[\hat{S}_{1}^{z}\otimes\hat{\mathbb{1}}_{2}\,
+\, \hat{\mathbb{1}}_{1}\otimes \hat{S}_{2}^{z}\,\r],\label{eq:HA}
\end{equation}
where $\hat{S}_{j}^z$, with $j=\{1,2\}$, denotes the operator that 
determines the energy levels of the detectors.
The operator $\hat{S}_{j}^{z}$ is defined as
\begin{equation}
\hat{S}_{j}^{z}
=\f{1}{2}\,\l(\vert e_{j}\rangle\, \langle e_{j}\vert
-\vert g_{j}\rangle\, \langle g_{j}\vert\r),
\end{equation}
where $\vert g_{j}\rangle$ and $\vert e_{j}\rangle$ represent the ground 
and excited states of the $j$-$\mathrm{th}$ atom. 
Moreover, note that, in the Hamiltonian~\eqref{eq:HA} describing the detectors, 
the quantity~$\hat{\mathbb{1}}$ represents the identity operator, and $\omega_{0}$
represents the transition energy corresponding 
to the collective two detector system. 
In particular, for identical, static detectors, the energy eigen states and 
eigen values for the two-atom system are given by~\cite{Arias:2015moa}
\begin{subequations}
\begin{eqnarray}
E_{e} &=& \omega_{0},~~~~\vert e\rangle = \vert e_{1} \rangle\,
\vert e_{2}\rangle,\quad\\
E_{s} &=& 0,~~~~~~\vert s\rangle = \f{1}{\sqrt{2}}\, \l(\vert e_{1}\rangle\, 
\vert g_{2}\rangle + \vert g_{1}\rangle\, \vert e_{2}\rangle\r),\\
E_{a} &=& 0,~~~~~~\vert a\rangle = \f{1}{\sqrt{2}}\,\l(\vert e_{1}\rangle\, 
\vert g_{2}\rangle - \vert g_{1}\rangle\, \vert e_{2}\rangle\r),\quad\\
E_{g} &=& -\omega_{0},~~\vert g\rangle = \vert g_{1}\rangle \vert g_{2}\rangle,
\end{eqnarray}
\end{subequations}
where $\vert g\rangle$ and $\vert e\rangle$ correspond to the ground and 
the excited states of the collective system, while $\vert s\rangle$ and 
$\vert a \rangle$ denote the symmetric and anti-symmetric maximally entangled 
Bell states. 
A pictorial representation of the different states and the associated 
energy levels of the two entangled detectors is illustrated in 
Fig.~\ref{fig:Energy-levels}. 

In $(2+1)$-dimensional Minkowski spacetime, the Hamiltonian of the massless, 
free scalar field is given by
\begin{equation}\label{eq:Hamiltonian-field}
H_\mathrm{F} = \frac{1}{2} \int \d^2{\bm x}\,
\l[\dot\Phi^2(\tx)+\l\vert\bm{\nabla}\Phi(\tx)\r\vert^2\r],
\end{equation}
where the overdot denotes differentiation with respect to the time 
coordinate, and $\bm{\nabla}$ denotes the spatial 
gradient.
The interaction Hamiltonian describing the monopole detectors and the scalar 
field is assumed to be 
\begin{equation}\label{eq:Hamiltonian-int}
H_\mathrm{I}
= \sum_{j=1}^{2}\mu_{j}\, m_j(\tau_{j})\,\kappa_{j}(\tau_{j})\,
\Phi[\tx_j(\tau_{j})],
\end{equation}
where $\mu_{j}$ denotes the strengths of the individual coupling between 
the detectors and the scalar field, while $m_{j}(\tau_{j})$ and 
$\kappa_{j}(\tau_{j})$ denote the monopole operators of the detectors
and the switching functions, respectively.
For identical atomic detectors, the coupling strengths between the detectors 
and the scalar field can be assumed to be the same, i.e. we can set $\mu_{1}
=\mu_{2}=\mu$. 
In such a case, the time evolution operator can be expressed as 
\begin{eqnarray}\label{eq:TimeEvolution-int}
\hat{U} &=& \mathcal{T}\; \exp\,\biggl\{-i\,\mu \int_{-\infty}^{\infty}\,
\biggl[\hat{m}_1(\tau_{1})\, \kappa_{1}(\tau_{1})\,
\hat{\Phi}[\tx_{1}(\tau_{1})]\,\d\tau_{1}\nn\\ 
& & +\,  \hat{m}_2(\tau_{2})\,\kappa_{2}(\tau_{2})\, 
\hat{\Phi}[\tx_{2}(\tau_{2})]\,\d\tau_{2} \biggr]\biggr\},
\end{eqnarray}
where $\mathcal{T}$ implies time ordering. 
Let $\vert \omega\rangle$ be the collective initial state of the two detector 
system, and $\vert \bar{\omega}\rangle$ be the collective final state. 
Also, let the initial state of the scalar field be the Minkowski vacuum $\vert 
\om \rangle$, and let $\vert \Theta\rangle$ be the final state of the scalar 
field.
Under these conditions, the transition amplitude from the initial state 
$\vert \omega, \om\rangle$ to the final state $\vert\bar{\omega},\Theta\rangle$ 
at the first order (when expanded in the strength of the coupling constant $\mu$) 
in perturbation theory is given by
\begin{eqnarray}\label{eq:Transition-amp}
\mathcal{A}_{\vert\omega,\om\rangle\to\vert\bar{\omega},\Theta\rangle} 
\!\! &=&\!\! \langle\Theta, \bar{\omega}\vert\hat{U}\vert \omega,\om\rangle\nn\\
& \simeq & -i\,\mu\,\langle\Theta,\bar{\omega}\vert\int_{-\infty}^{\infty} \,
\biggl[\kappa_1\,\hat{m}_1\,\hat{\Phi}(\tx_{1})\, \d\tau_{1} \nn\\
& & +\, \kappa_{2}\,\hat{m}_2\, \hat{\Phi}(\tx_{2})\, \d\tau_{2}\biggr]
\vert\omega,\om\rangle.
\end{eqnarray}
Note that we shall be interested in examining the final state of the detectors.
The total transition probability of the detectors can be arrived at from the 
above transition amplitude by summing over all the final states 
$\l\{\vert\Theta\rangle\r\}$ of the field.
The total transition probability of the two detectors can be expressed as 
\begin{eqnarray}\label{eq:Transition-prob}
\Gamma_{\vert \omega\rangle\to\vert\bar{\omega}\rangle}(\cE) 
&=& \sum_{\l\{\vert\Theta\rangle\r\}} \mathcal{A}_{\vert\omega,\om\rangle
\to \vert\bar{\omega},\Theta\rangle}\, \mathcal{A}_{\vert \omega,\om\rangle
\to\vert\bar{\omega},\Theta\rangle}^{\ast}\nn\\
&\simeq & \mu^2 \sum_{j,l=1}^{2} 
m_{j}^{\bar{\omega}\omega\ast}\, 
m_{l}^{\bar{\omega}\omega}\,F_{jl}(\cE),
\end{eqnarray}
where $\cE=E_{\bar{\omega}}-E_{\omega}$, with $E_{\omega}$ and
$E_{\bar{\omega}}$ denoting the energy eigen values associated with 
the states $\vert \omega\rangle$ and $\vert \bar{\omega}\rangle$, and 
$m_{j}^{\bar{\omega}\omega} 
= \langle \bar{\omega} \vert \hat{m}_{j}(0)\vert\omega\rangle$. 
As we shall discuss below, the quantities $F_{jl}(\cE)$---which we shall
refer to as the auto or the cross transition probabilities---depend on 
the trajectory of the detectors.

Meanwhile, let us understand the values that the 
quantity~$m_{j}^{\bar{\omega}\omega}$ can take.
We shall assume that the operator describing the monopole moment of the
detectors is given by
\begin{equation}
\hat{m}_{j}(0)=\vert e_{j}\rangle \langle g_{j}\vert
+ \vert g_{j}\rangle \langle e_{j}\vert.\label{eq:mo}
\end{equation} 
This expression for the monopole operators can be utilized to determine the 
contributions due to specific transitions between the collective initial and 
final states of the two detectors.
For instance, it can be shown that the transition from the collective ground 
state~$\vert g \rangle$ to the collective excited state~$\vert e\rangle$ (or 
the other way around) of the entangled detectors is not possible since 
$m_{j}^{ge} = m_{j}^{eg} =0$. 
Also, one finds that $m_{1}^{se} = m_2^{se} =1/\sqrt{2}$ and $m_{1}^{ae} 
= -m_{2}^{ae} = -1/\sqrt{2}$, which denote the amplitudes for transitions 
between the symmetric and anti-symmetric Bell states (viz. $\vert 
s \rangle$ and $\vert a \rangle$) and the excited state of the detectors, 
respectively.
Moreover, one can show that the amplitudes for the transitions from the 
collective ground state to the symmetric and anti-symmetric Bell states are 
given by $m_{1}^{gs}=m_{2}^{gs}=1/\sqrt{2}$ and $m_{1}^{ga}=-m_{2}^{ga}=1/\sqrt{2}$.
The energy levels associated with the different states of the two entangled
detectors and the various possible transitions are illustrated diagrammatically 
in Fig.~\ref{fig:Energy-levels} (taken from Ref.~\cite{Costa:2020aqa}).

Let us now shift our attention to the transition probabilities 
$F_{jl}(\cE)$ in Eq.~\eqref{eq:Transition-prob}.
The explicit form of the transition probabilities $F_{jl}(\cE)$ are found
to be
\begin{eqnarray}\label{eq:Transition-coeff}
F_{jl}(\cE) 
&=& \int_{-\infty}^{\infty}\d\tau_l'  \int_{-\infty}^{\infty}\d\tau_j\, 
\mathrm{e}^{-i\,\cE\,(\tau_{j}-\tau_{l}')}\nn\\
& &\times\, G_{jl}^{+}[\tx_{j}(\tau_j),\tx_{l}(\tau_l')]\, 
\kappa_j(\tau_j)\,\kappa_l(\tau_l'),
\end{eqnarray}
where the quantity $G_{jl}^{+}[\tx_{j}(\tau_j),\tx_{l}(\tau_l')]$ denotes the 
positive frequency Wightman function evaluated along the trajectories of the 
detectors.
The positive frequency Wightman function is defined as 
\begin{equation}\label{eq:Two-point-fn-gen}
G_{jl}^{+}[\tx_{j}(\tau_j),\tx_{l}(\tau_l')] 
= \langle \om \vert \hat{\Phi}[\tx_{j}(\tau_{j})]\,
\hat{\Phi}[\tx_{l}(\tau_{l}')]\vert \om \rangle.
\end{equation}
In the following sections, we shall evaluate the transition probabilities of 
detectors that are in motion on circular trajectories in $(2+1)$-dimensional
Minkowski spacetime.
We shall evaluate the responses of the detectors in the Minkowski vacuum as 
well as in a thermal bath.  
As we shall see, it proves to be convenient to work in terms of the polar 
coordinates to arrive at the Wightman function along the trajectories of 
the detectors when they are in circular motion.


\section{Response of a detector in circular motion}\label{sec:rrd-mv-tb}

In this section, we shall derive the response of a {\it single}\/ 
Unruh-DeWitt detector that is interacting with a scalar field.
The results we obtain in this situation will prove to be helpful
for understanding the results in the case of the two entangled 
detectors.
Consider a massless and minimally coupled scalar field~$\Phi$ that
is described by the action
\begin{equation}\label{eq:action-scalar-field}
S[\Phi] = -\int \d^3 x\,  \sqrt{-g}\,
\f{1}{2}\,g^{\mu\nu}\,\pa_{\mu}\Phi\,\pa_{\nu}\Phi.
\end{equation}
On varying the action, we can obtain the equation of motion of the scalar 
field to be
\begin{equation}
\Box\Phi=\f{1}{\sqrt{-g}}\,
\pa_{\mu}\l(\sqrt{-g}\,g^{\mu\nu}\,\pa_{\nu}\r)\Phi=0.
\end{equation}

In order to examine the behavior of a rotating detector in Minkowski spacetime, 
it proves to be convenient to work in the polar coordinates so that, in
$(2+1)$-dimensions, the spacetime coordinates are given by $\tx=(t,\rho,\phi)$.
In these coordinates, the normal modes of the massless scalar field can be 
obtained to be
\begin{equation}\label{eq:modes-cylindrical}
u_{qm}(\tx) 
= \f{1}{\sqrt{4\,\pi}}\, \mathrm{e}^{-i\,q\,t}\,
J_{m}(q\,\rho)\, \mathrm{e}^{i\,m\,\phi},
\end{equation}
where $0 \leq q <\infty$, $m$ is an integer, and $J_n(z)$ denotes the Bessel 
function of order~$n$.
On quantization, the scalar field can be decomposed in terms of the above
normal modes $u_{qm}(\tx)$ as follows:
\begin{equation}\label{eq:scalar-field-decomposition}
\hat{\Phi}(\tx) = \int_{0}^{\infty} \d q \sum_{m=-\infty}^{\infty} 
\l[\hat{a}_{qm}\,u_{qm}(\tx) + \hat{a}_{qm}^\dag\,u_{qm}^\ast(\tx) \r],
\end{equation}
where $\hat{a}_{qm}$ and $\hat{a}_{qm}^\dag$ are the creation and the 
annihilation operators which satisfy the following standard commutation
relations: 
\begin{subequations}
\begin{eqnarray}
\l[\hat{a}_{qm}, \hat{a}_{q'm'}\r] 
&=& \l[\hat{a}_{qm}^\dag, \hat{a}_{q'm'}^{\dag}\r] = 0,\\
\l[\hat{a}_{qm}, \hat{a}_{q'm'}^{\dag}\r] &=& \delta^{(1)}(q-q')\,\delta_{mm'}.
\end{eqnarray}
\end{subequations}

In the case of a single detector, the transition 
probability~\eqref{eq:Transition-coeff} simplifies to be
\begin{eqnarray}\label{eq:F}
F(\cE) &=& \int_{-\infty}^{\infty}\d\tau'  \int_{-\infty}^{\infty}\d\tau\, 
\mathrm{e}^{-i\,\cE\,(\tau-\tau')}\nn\\
& &\times\,G^{+}[\tx(\tau),\tx(\tau')]\,\kappa(\tau)\,\kappa(\tau').
\end{eqnarray}
In our discussion, we shall be interested in examining the response of detectors 
that are moving on circular trajectories.
Also, we shall evaluate the response of the detectors in the Minkowski vacuum 
and in a thermal bath.
We shall utilize the mode functions~\eqref{eq:modes-cylindrical} in the 
polar coordinates to arrive at the Wightman function $G^+(\tx,\tx')$ in 
both these situations. 
One finds that, in these situations, the Wightman function along the trajectory 
of a detector in circular motion is invariant under the time translation in 
the proper time in the frame of the detector, i.e. $G^+[\tx(\tau),\tx(\tau')]
=G^+(\tau,\tau')=G^+(u)$, where $u=(\tau-\tau')$.
It is well known that such a time translation invariance allows one to define 
the transition probability rate of the detector.
Often, in these contexts, the Wightman function is first evaluated by 
summing over all the normal modes of the quantum field, 
before evaluating the transition probability rate of the detector.
As we shall see, to arrive at the transition probability rate of the rotating
detector, rather than explicitly evaluate the Wightman function, it proves to 
be convenient to first carry out the integral over the quantity~$u$, and then 
sum over the modes.
In the following subsections, we shall evaluate the transition probability 
rate of a rotating detector that has been switched on for infinite as well
as a finite time interval.


\subsection{Detector switched on for infinite duration}

Let us first consider the situation wherein the detector remains switched 
on for all times.
In such a situation, the switching function~$\kappa(\tau)$ reduces to 
unity.


\subsubsection{Response in the Minkowski vacuum}

Let us now evaluate the response of the rotating detector in the Minkowski
vacuum.
The Wightman function $G^+(\tx,\tx')$ associated with the massless scalar 
field in the Minkowski vacuum can be expressed as (in this
regard, also see App.~\ref{app:Evln-GreensFn})
\begin{eqnarray}\label{eq:Greenfn-1detector}
G^+(x,x') &=& \langle \om\vert \hat{\Phi}(\tx)\,\hat{\Phi}(\tx')\vert \om\rangle\nn\\ 
&=& \int_{0}^{\infty} \f{\d q}{4\,\pi}\,  \sum_{m=-\infty}^{\infty} 
J_{m}(q\,\rho)\,J_{m}(q\,\rho')\nn\\ 
& &\times\,\mathrm{e}^{-i\,q\,(t-t')}\, \mathrm{e}^{i\,m\,(\phi-\phi')},
\end{eqnarray}
where we have made use of the fact that the Bessel
functions~$J_n(x)$ are real for integer values of~$n$ and real arguments.
In order to arrive at the response of the detector, we need to calculate the 
above Wightman function along the trajectory of the detector.
Let us assume that the detector is moving along a circular trajectory with 
radius~$\sigma$, at a constant angular velocity~$\Omega$.
If $\tau$ is the proper time in the frame of the detector, then the 
trajectory of the detector is given by (see, for example, 
Refs.~\cite{Gutti:2010nv, Costa:2020aqa}) 
\begin{equation}\label{eq:CoordTrans-rotating}
t=\gamma\,\tau,\quad \rho=\sigma,\quad \phi=\gamma\,\Omega\,\tau,
\end{equation}
where $\gamma=1/\sqrt{1-v^2}$ is the Lorentz factor associated with the 
linear velocity $v=\sigma\,\Omega$ of the detector.
Along such a trajectory, the Wightman function above reduces the following 
form:
\begin{equation}
G^+(u)= \int_{0}^{\infty} \f{\d q}{4\,\pi}\, \sum_{m=-\infty}^{\infty} 
J_{m}^2(q\,\sigma)\,\mathrm{e}^{-i\,\gamma\,(q-m\,\Omega)\,u},\label{eq:wfn-rt}
\end{equation}
where, as we mentioned above, $u=(\tau-\tau')$.
In such a situation, we can define the transition probability rate of the 
detector to be
\begin{equation}
R(\cE) = \int_{-\infty}^{\infty} \d u\, 
\mathrm{e}^{-i\,\cE\,u}\, G^+(u).\label{eq:tpr}
\end{equation}

On substituting the Wightman function~\eqref{eq:wfn-rt} along the trajectory 
of the rotating detector in the above expression, we find that the transition
probability rate can be expressed as
\begin{eqnarray}\label{eq:response-fn}
R(\cEb) &=& \sum_{m=-\infty}^{\infty}\,
\int_{0}^{\infty} \f{\d q}{2\,\gamma}\, 
J^2_{m}(q\,\sigma)\,\delta^{(1)}[q-(m-\cEb)\,\Omega]\nn\\
&=& \f{1}{2\,\gamma}\,\sum_{m\ge \cEb}^{\infty} 
J^2_{m}\l[(m-\cEb)\,v\r],
\end{eqnarray}
where we have introduced the dimensionless energy gap $\cEb=\cE/(\gamma\,\Omega)$ 
and, recall that, $v=\sigma\, \Omega$ is the linear velocity of the detector.
Note that, since $q\in[0,\infty)$, the Dirac delta function in the above
expression leads to non-zero contributions only when $m\ge \cEb$, which 
is reflected in the lower limit of the sum in the final expression.

\begin{figure*}
\centering
\includegraphics[width=0.475\linewidth]{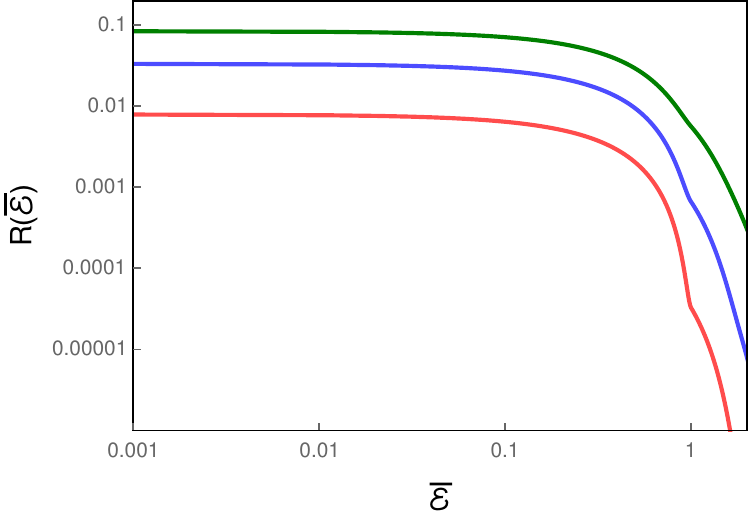}
\hskip 10pt
\includegraphics[width=0.475\linewidth]{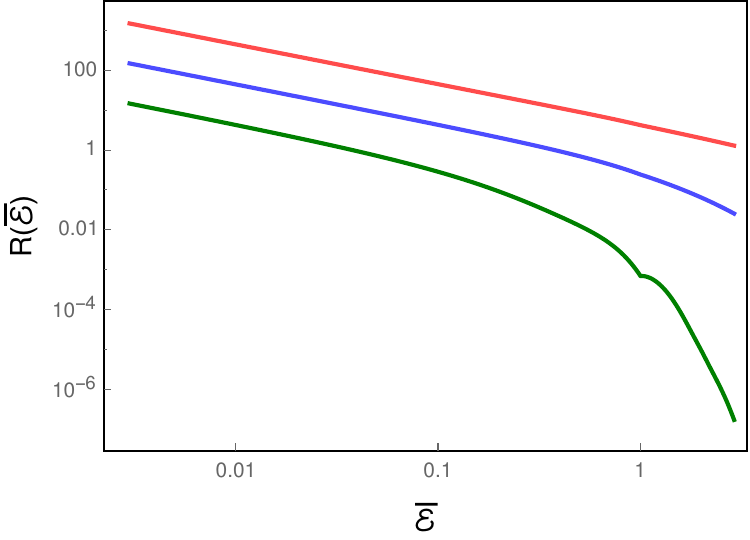}
\caption{The transition probability rate~$R(\cEb)$ of the Unruh-DeWitt detector
that is moving on a circular trajectory and remains switched on forever has been 
plotted as a function of the dimensionless energy gap~$\cEb$.
We have plotted the response of the detector in the Minkowski vacuum (on the 
left) and in a thermal bath (on the right). 
In the case of the Minkowski vacuum, we have plotted the results for three
different values of the velocity of the detector, viz. $v=(0.25, 0.5, 0.75)$ 
(in red, blue and green) and, in the case of the thermal bath, we have fixed 
the velocity to be $v=0.5$ and have plotted the results for three different
values of the dimensionless inverse temperature, viz. $\bar{\beta}=(0.1,1,10)$ 
(in red, blue and green).
We have arrived at these results by summing until $(m-\cEb) = 50$ [in 
Eqs.~\eqref{eq:response-fn} and~\eqref{eq:response-fn-infinite-thermal}]
and we have confirmed that summing up to higher values of $m$ does not 
significantly alter the results.
Note that the transition probability rate of the rotating detector is larger
at smaller energies for a given velocity and temperature.
Also, we find that, for a given energy and temperature, the rate
is higher at a higher velocity of the detector.
Moreover, for a given energy and velocity, the transition probability rate is 
higher when the temperature of the bath is higher. 
Clearly, this can be attributed to the fact that there are more quanta available 
to excite the detector at higher temperatures.}\label{fig:srd}
\end{figure*}

Evidently, the transition probability rate~\eqref{eq:response-fn} of the 
detector in circular motion depends only on its linear velocity~$v$.
The sum in the expression for the transition probability rate proves to be 
difficult to evaluate analytically.
But, it converges quickly enough to be computed numerically.
For a given value of~$\cEb$, we find that the sum converges exponentially
beyond a certain value of~$m$.
For instance, we find that, when $\cEb$ is chosen to be unity, for $v=(0.25,0.5,
0.75)$, the quantity $J^2_{m}\l[(m-\cEb)\,v\r]$ that appears in the sum 
characterizing the transition probability rate of the detector decreases at least 
as fast as $(\mathrm{e}^{-2.0\,m},\mathrm{e}^{-0.9\,m},\mathrm{e}^{-0.25\,m})$, 
respectively, at suitably large~$m$.
We should clarify that we have confirmed this behavior for adequately large 
values of~$m$, much beyond the values we sum over to arrive at the result.
The rapid convergence of the sum allows us to easily calculate the transition 
probability rate of the rotating detector numerically by summing up to a
finite value of $m$~(in this regard, also see, App.~\ref{app:m}).
We have also ensured that, over the domain of $\cEb$ we focus on, the contributions
beyond the maximum value of~$m$ we have worked with are insignificant.
In Fig.~\ref{fig:srd}, we have illustrated the transition probability rate of the 
detector as a function of the dimensionless energy gap $\cEb$ for a few different 
values of the velocity.
The figure suggests that, higher the velocity of the detector, higher is its 
transition probability rate.
Moreover, for a given velocity, the transition probability of the detector is
larger at smaller values of the energy gap such that~$\cEb\ll 1$.


\subsubsection{Response in a thermal bath}

Let us now turn to evaluate the response of the detector in a thermal bath.
We shall assume that the massless scalar field $\Phi$ of our interest 
is in equilibrium with a thermal bath maintained at the inverse 
temperature~$\beta$. 
We can utilize the decomposition~\eqref{eq:scalar-field-decomposition} of 
the scalar field in terms of the normal modes~\eqref{eq:modes-cylindrical} 
to arrive at the following expression for the Wightman function at a finite 
temperature (in this context, also see App.~\ref{app:Evln-GreensFn}):
\begin{eqnarray}\label{eq:Greenfn-2p1-thermal}
G^{+}_{\beta}(\tx,\tx')
&=& \int_{0}^{\infty}\frac{\d q}{4\,\pi}\, \sum_{m=-\infty}^{\infty}
J_{m}(q\,\rho)\,J_{m}(q\,\rho')\nn\\
& &\times\,\biggl\{\f{\mathrm{e}^{-i\,[q\,(t-t')- m\,(\phi-\phi')]}}{1
-\mathrm{e}^{-\beta\, q}}\nn\\ 
& &+\,\f{\mathrm{e}^{i\,[q\,(t-t')- m\,(\phi-\phi')]}}{\mathrm{e}^{\beta\,q}
-1}\biggr\}.
\end{eqnarray}
Along the trajectory~\eqref{eq:CoordTrans-rotating} of the rotating detector, 
this finite temperature Wightman function too turns out to be invariant 
under translations in the proper time of the detector as in the Minkowski 
vacuum.
We find that, in the frame of the rotating detector, the finite temperature
Wightman function reduces to 
\begin{eqnarray}\label{eq:Greenfn-2UDD-thermal-i}
G^{+}_{\beta}(u)
&=& \int_{0}^{\infty}\frac{\d q}{4\,\pi}\, \sum_{m=-\infty}^{\infty}
J^{2}_{m}(q\,\sigma)\nn\\ 
& &\times\,\l[\f{\mathrm{e}^{-i\,\gamma\, (q-m\,\Omega)\,u}}{1
-\mathrm{e}^{-\beta\, q}}
+\f{\mathrm{e}^{i\,\gamma\, (q-m\,\Omega)\,u}}{\mathrm{e}^{\beta\,q}-1}\r].
\qquad
\end{eqnarray}
Since the Wightman function depends only on the quantity~$u$, we can 
define a transition probability rate for the detector as in the Minkowski 
vacuum [cf. Eq.~\eqref{eq:tpr}].
The transition probability rate of the rotating detector at a finite 
temperature can be easily evaluated to be 
\begin{eqnarray}\label{eq:response-fn-infinite-thermal}
R(\cEb) 
&=& \sum_{m=-\infty}^{\infty}\, \int_{0}^{\infty} \f{\d q}{2\,\gamma}\, 
J^2_{m}(q\,\sigma)\,
\biggl\{\f{\delta^{(1)}[q-(m-\cEb)\,\Omega]}{1-\mathrm{e}^{-\beta\, q}}\nn\\
& &+\,\frac{\delta^{(1)}[q-(m+\cEb)\,\Omega]}{\mathrm{e}^{\beta\, q}-1}\biggr\}\nn\\
&=& \f{1}{2\,\gamma}\,\biggl\{\sum_{m\ge \cEb}^{\infty}\,
\f{J^2_{m}\l[(m-\cEb)\,v\r]}{1-\mathrm{e}^{-\bar{\beta}\,(m-\cEb)}}\nn\\
& &+\,\sum_{m\ge -\cEb}^{\infty} 
\frac{J^2_{m}\l[(m+\cEb)\,v\r]}{\mathrm{e}^{\bar{\beta}\,(m+\cEb)}-1}\biggr\},
\end{eqnarray}
where $\bar{\beta}=\beta\,\Omega$ denotes the dimensionless inverse temperature 
of the bath.
We had noted earlier that, the quantity $J^2_{m}\l[(m-\cEb)\,v\r]$ which appears 
in the sum characterizing the transition probability rate decreases exponentially
at large~$m$.
Though the sums in the above expression are again difficult to evaluate analytically,
such a rapid convergence allows us to compute them numerically rather easily 
(again, in this regard, see, App.~\ref{app:m}). 
Actually, in the case of the second term, the exponential in the denominator also
aids in a faster convergence of the sum. 
In Fig.~\ref{fig:srd}, we have plotted the above transition probability rate 
of the detector for a fixed value of the velocity $v$ and a few different 
values of the dimensionless inverse temperature~$\bar{\beta}$.
It should be clear from the figure that, larger the temperature (or, 
equivalently, smaller the value of $\bar{\beta}$), larger is the 
transition probability rate of the rotating detector. 

At this stage, there is a technical point that we need to discuss.
Actually, the finite temperature Wightman function~\eqref{eq:Greenfn-2UDD-thermal-i} 
contains an infrared divergence.
As $q\to 0$, the functions $J_m^2(q\,\sigma)$ behave as $q^{2\,m}$ and, hence, 
the $m=0$ term in the Wightman function diverges logarithmically in this limit 
{\it even for finite separation of the spacetime points}.\/
This behavior is a surprising and less known peculiarity of the thermal Green's 
function in $(2+1)$-dimensional Minkowski spacetime and, in fact, the divergence 
is absent at zero temperature [as can be easily checked with the Wightman 
function~\eqref{eq:wfn-rt}].
Also, it can be readily shown that such an infrared divergence is not encountered 
in $(3+1)$-dimensional Minkowski spacetime.
We should point out that the infrared divergence occurs in addition to the 
ultraviolet divergence which arises at large~$q$.
The ultraviolet divergence can, as usual, be regulated using the $(i\,\epsilon)$-prescription.  
Note that, in arriving at the rate $R(\cEb)$ in Eq.~\eqref{eq:response-fn-infinite-thermal},
we chose to calculate the integral over~$u$ first before evaluating the integral over~$q$.
In the process, the infrared divergence is transferred to the $m=\pm\cEb$ 
term in the sum, and it manifests itself only in the $\cEb \to 0$ limit.
But, since we have assumed that $\cE>0$, we do not actually encounter the divergence
when evaluating the sum.
In the following section, when we consider detectors which are switched on for 
a finite duration, we shall find that the divergence at the finite temperature 
cannot be circumvented in a similar manner.
To handle the divergence, we shall adopt a procedure which allows us to reproduce 
the results in the different limits, viz. at zero temperature and when the detector 
is switched on for infinite duration.

There is another point that we need to clarify regarding the results illustrated 
in Fig.~\ref{fig:srd}.
Note that, in the case of results plotted at a finite temperature, the transition 
probability rate~$R(\cEb)$ turns out to be more than unity for small values of~$\cEb$. 
This may cause concern.
But, it occurs due to the fact that we have dropped an overall factor of~$\vert\mu\vert^2$, 
where $\mu$ is the coupling constant
that determines the strength of the interaction
between the detector and the field [cf. Eq.~\eqref{eq:Hamiltonian-int}], when 
evaluating the transition probability rate.
In order for the perturbative expansion of the time evolution operator in 
Eq.~\eqref{eq:TimeEvolution-int} to be valid, we require $\mu$ to be much 
smaller than unity. 
Evidently, for a suitably small value of $\vert \mu\vert^2$, the transition 
probability rate will reduce to a value less than unity for all energies~$\cEb$.

We will now show that the transition probability rate of the rotating detector in 
a thermal bath we have obtained above corresponds to the accumulation of different
types of radiative processes that occur in the system. 
Let us introduce the quantities
\begin{equation}
\mathcal{N}_{r}(q,\cEb)
= \f{1}{2\,\gamma}\sum_{m=-\infty}^{\infty} J^2_{m}(q\,\sigma)\,
\delta^{(1)}[q-(m-\cEb)\,\Omega]
\end{equation}
and 
\begin{equation}\label{eq:response-fn-infinite-thermal3}
\mathcal{N}_{\beta}(q)=\frac{1}{\mathrm{e}^{\beta\,q}-1}.
\end{equation}
Since $(1-\mathrm{e}^{-x})^{-1}=1+(\mathrm{e}^{x}-1)^{-1}$, in terms of 
the above quantities, we can re-express the first equality of
Eq.~\eqref{eq:response-fn-infinite-thermal} in the following form:
\begin{eqnarray}\label{eq:response-fn-infinite-thermal2}
R(\cEb)
&=& \underbrace{\int_{0}^{\infty} \d q\,\mathcal{N}_{r}(q,\cEb)}_{R_1(\cEb)} \nn\\
& &+\, \underbrace{\int_{0}^{\infty} \d q\,\mathcal{N}_{\beta}(q)\,
\l[\mathcal{N}_{r}(q,\cEb)+\mathcal{N}_{r}(q,-\cEb)\r]}_{R_2(\cEb,\beta)}.\qquad
\end{eqnarray}
In other words, we can express the transition probability rate of the rotating 
detector in a thermal bath as a sum of the two contributions $R_1(\cEb)$ and 
$R_2(\cEb,\beta)$, which, as we shall soon discuss, can be attributed to 
different types of radiative processes. 
Note that the first term $R_1(\cEb)$ is only a function of~$\cEb$ and is 
independent of~$\beta$, whereas the second term $R_2(\cEb,\beta)$ is a 
function of both~$\cEb$ and~$\beta$. 
The fact that the term $R_1(\cEb)$ is the same as the first equality in 
Eq.~\eqref{eq:response-fn} clearly suggests that it corresponds to the 
response of the rotating detector in the Minkowski vacuum.
The contribution arises due to modes with the magnetic quantum numbers $m_1 = 
(q/\Omega)+\cEb$ for a given value of momentum~$q$. 
A contribution from the Minkowski vacuum can always be expected to occur and the
contribution can be interpreted as arising due to the {\it spontaneous excitation}\/
of the detector.
The interesting aspect of the term $R_2(\cEb,\beta)$ is the appearance of the 
factor $\mathcal{N}_{\beta}(q)$.
The factor represents the distribution of scalar field modes with momentum~$q$
and it reflects the fact that the scalar field is immersed in a thermal bath.
Actually, the contribution $R_2(\cEb,\beta)$ consists of two parts. 
The first part involving $\mathcal{N}_{r}(q,\cEb)$ can be interpreted as the 
excitation of the rotating detector due to the thermal nature of the scalar field, 
with modes corresponding to the magnetic quantum number~$m_1$ contributing to
the transition probability rate of the detector. 
The second part involving $\mathcal{N}_{r}(q,-\cEb)$ too signifies the excitation 
of detector by the thermal character of the field, but with the contributions
arising from modes with a different set of magnetic quantum number, viz. 
$\bar{m}_1=(q/\Omega)-\cEb$. 
Evidently, these latter two contributions can be attributed to the combined 
effects of both the circular motion of the detector as well as the thermal
bath.
Since these contributions are influenced by the presence of the thermal bath 
and vanish in its absence (i.e. when $\beta\to\infty$), these contributions
can be interpreted as arising due to the {\it stimulated excitation}\/ of the 
detector.
Therefore, the overall response of the rotating detector in the thermal bath
can be interpreted as arising due to the accumulation of three types of 
radiative process---two processes (spontaneous and stimulated excitation) to 
which the scalar field modes with the magnetic quantum number~$m_1$ 
contribute and another process (stimulated excitation) which arises due to the
contributions by the modes with quantum number~$\bar{m}_1$.  

In this regard, it may be pointed out that a similar interpretation has also 
been suggested earlier for the response of a uniformly accelerated detector in
a thermal bath (see Refs.~\cite{Kolekar:2013aka,Kolekar:2013xua,Kolekar:2013hra}). 
However, we observe a noticeable difference between the transition probability
rates of accelerated detectors and our present system. 
In the accelerated case, the second term in Eq.~\eqref{eq:response-fn-infinite-thermal2},
which we had interpreted as due to stimulated excitation, has a simpler structure. 
It is composed of a purely thermal contribution and an excitation due to the 
acceleration effects stimulated by the thermal bath. 
Therefore, for an accelerated detector in a thermal bath, there are two 
independent, spontaneous excitations---one is due to the acceleration (known 
as the Unruh effect), and the other is purely due to the thermal bath; apart 
from stimulated excitation, which is influenced by the thermal background. 
Whereas, in the present study, we do not find any contribution due to the 
thermal bath that is independent of rotation. 
This aspect of the transition probability rates of rotating detectors in a
thermal bath provides a signature distinct from the case of accelerated 
detectors.


\subsection{Detector switched on for a finite duration}

Let us now consider the case wherein the detectors are assumed to be 
switched on for a finite time interval, say,~$T$.
We shall consider the switching functions $\kappa(\tau)$ to be of the 
following Gaussian form~\cite{Sriramkumar:1994pb}:
\begin{equation}
\kappa(\tau)=\mathrm{exp}\,\l(-\f{\tau^2}{T^2}\r),\label{eq:gsf}
\end{equation}
where, evidently, $T$ denotes the duration for which the detector remains
effectively switched on.
In the presence of such a switching function, the transition probability
of the detector is given by [cf. Eq.~\eqref{eq:Transition-coeff}]
\begin{eqnarray}
F_T(\cE) 
&=& \int_{-\infty}^{\infty}\d\tau^{\prime} 
\int_{-\infty}^{\infty}\d\tau\, 
\mathrm{e}^{-i\,\cE\,(\tau-\tau')}\, G^{+}(\tau,\tau')\nn\\
& &\times\,\mathrm{exp}\,\l[-(\tau^2+{\tau^{\prime}}^2)/T^2\r].
\end{eqnarray}
On changing variables to $u=(\tau-\tau')$ and  $v=(\tau+\tau')$, the
integral can be expressed as
\begin{eqnarray}
F_T(\cE) 
&=& \int_{-\infty}^{\infty}\f{\d v}{2}\, \mathrm{e}^{-v^2/(2\,T^2)}\nn\\
& &\times\, \int_{-\infty}^{\infty}\d u\, \mathrm{e}^{-i\,\cE\,u}\,
G^{+}(u,v)\, \mathrm{e}^{-u^2/(2\,T^2)}.\qquad
\end{eqnarray}
In cases wherein the Wightman function is invariant under time translations
in the frame of the detector, i.e. when $G^{+}(\tau,\tau')=G^{+}(u)$---as 
in the case of the rotating detector [cf. Eqs.~\eqref{eq:wfn-rt} 
and~\eqref{eq:Greenfn-2UDD-thermal-i}]---we can carry out the Gaussian 
integral over~$v$ to define the transition probability rate of the detector 
as follows:
\begin{eqnarray}
R_T(\cE) &=& \f{F_T(\cE)}{\sqrt{(\pi/2)}\;T}\nn\\
&=&\int_{-\infty}^{\infty}\d u\, \mathrm{e}^{-i\,\cE\,u}\, 
G^{+}(u)\, \mathrm{e}^{-u^2/(2\,T^2)}.\qquad\label{eq:tprT}
\end{eqnarray}
Note that, as required, $R_T(\cE) \to R(\cE)$ 
[cf. Eq.~\eqref{eq:tpr}] when $T\to \infty$.
We shall now utilize the above expression to evaluate the finite time 
response of the rotating detector in the Minkowski vacuum and in a 
thermal bath.


\subsubsection{Response in the Minkowski vacuum}

\begin{figure*}
\centering
\includegraphics[width=0.475\linewidth]{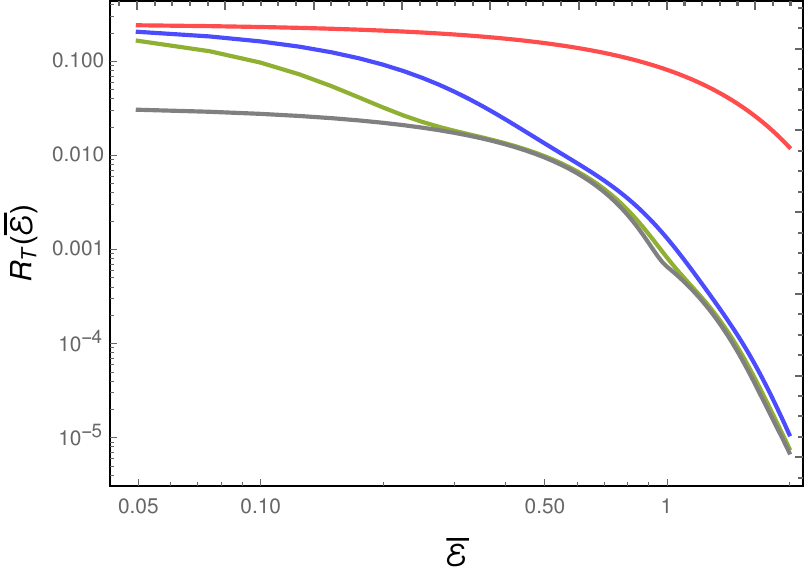}
\hskip 10pt
\includegraphics[width=0.475\linewidth]{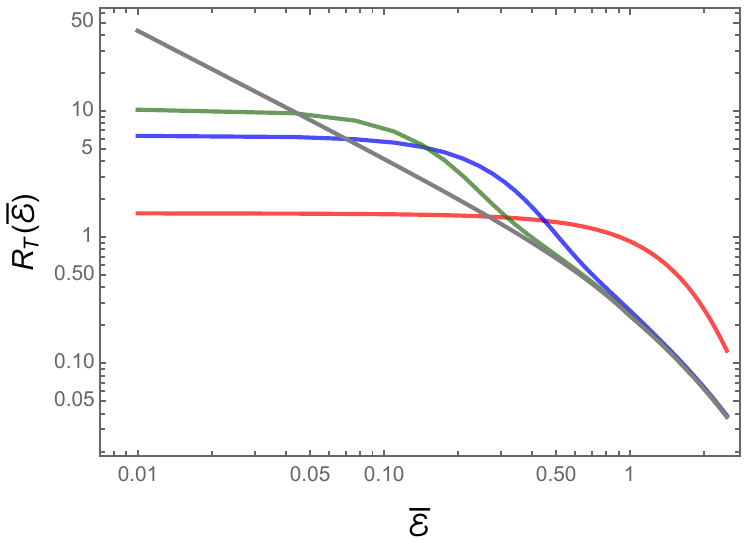}
\caption{The transition probability rate $R_{T}(\cEb)$ of the detector in 
motion on a circular trajectory that has been switched on for a finite
time~$T$ has been plotted as a function of the dimensionless energy gap~$\cEb$.
We have plotted the results for $\bar{T}=(1,5,10)$ (in red, blue and green),
assuming the field is in the Minkowski vacuum (on the left) and in a thermal 
bath (on the right).
We have set $v=0.5$ and $\bar{\beta}=1$ in the plotting of these figures.
We have arrived at the results for a finite time by integrating over $x$ from 
zero up to $10^2$, and carrying the sum over $m$ until $(m - \cEb) = 10$.
We have checked that increasing the upper limits of the integral and
the sum does not significantly change the results we have obtained.
In the figures, we have also indicated the results in the case wherein 
the detector is switched on forever, i.e. when $\bar{T}\to \infty$ (in 
gray).
Interestingly, we find that the transition probability rate of the detector
in the Minkowski vacuum is higher when it is switched on for a shorter duration. 
We find that, in a thermal bath, the transition probability rate of the detector 
also exhibits a similar behavior at high~$\cEb$ (when $\cEb \gtrsim 0.5$, for
the values of the parameters we have worked with), while at low $\cEb$ the 
behavior is reversed. 
At a sufficiently low temperature (say, $\bar{\beta} \gtrsim 150$), when $\bar{T}$ 
is decreased, we observe that the transition probability rate increases over the 
whole domain of $\cEb$, just like in the case of the Minkowski vacuum (in this 
context, see Fig.~\ref{fig:R11-fT-Th-Vbeta}).}\label{fig:RT11-fT-Mink-Vv}
\end{figure*}

On utilizing the expression~\eqref{eq:wfn-rt} for the Wightman function
in the Minkowski vacuum along the circular trajectory, the transition 
probability rate of the detector that is switched on for a finite time
interval~$T$ can be expressed as
\begin{eqnarray}
R_T(\cE) 
&=& \int_{0}^{\infty} \f{\d q}{4\,\pi}\, 
\sum_{m=-\infty}^{\infty} J^2_{m}(q\,\sigma)\nn\\
& &\times\,\int_{-\infty}^{\infty}\d u\, 
\mathrm{e}^{-i\,\l[\cE+\gamma\,(q-m\,\Omega)\r]\,u}\, 
\mathrm{e}^{-u^2/(2\,T^2)}.\qquad\;\;
\end{eqnarray}
The Gaussian integral over~$u$ can be calculated easily to arrive at
\begin{eqnarray}\label{eq:RT-single-detector}
R_T(\cE) 
&=& \sqrt{2\,\pi}\;T\, \sum_{m=-\infty}^{\infty}\, 
\int_{0}^{\infty} \f{\d q}{4\,\pi}\, 
J^2_{m}(q\,\sigma)\nn\\
& &\times\,\mathrm{e}^{-\l[\cE+\gamma\,(q-m\,\Omega)\r]^2\,T^2/2}.\qquad
\end{eqnarray}
Recall that the Dirac delta function can be represented in terms of the 
Gaussian function as follows:
\begin{equation}
\delta^{(1)}(z) = 
\lim_{\alpha \to \infty} 
\f{\alpha}{\sqrt{2\,\pi}}\, \mathrm{e}^{-\alpha^2\,z^2/2}.
\end{equation}
Therefore, in the limit $T\to \infty$, the 
expression~(\ref{eq:RT-single-detector}) for the finite time response 
rate of the detector reduces to the result~(\ref{eq:response-fn}) for 
detectors that remain switched on for infinite time (with the 
identification of $\alpha=\gamma\, T$), as required. 
Upon noting that $J_n^2(x)=J_{-n}^2(x)$ for integer~$n$ and real~$x$ and
setting $q=x\,\Omega$, the integral~(\ref{eq:RT-single-detector}) can be 
expressed as
\begin{widetext}
\begin{eqnarray}\label{eq:response-fn-finite-Minkowski}
R_T(\cEb) 
&=& \f{\sqrt{2\,\pi}\;\bar{T}}{4\,\pi\gamma}\,
\mathrm{e}^{-\cEb^2\,\bar{T}^2/2}\,\Biggl\{\int_{0}^{\infty} \d x\,
J^2_{0}(x\,v)\,
\mathrm{e}^{-\l[(x^2/2)+x\,\cEb\r]\,\bar{T}^2}\,\nn\\
& &+\,2\,\sum_{m=1}^{\infty}\,  \mathrm{e}^{-m^2\,\bar{T}^2/2}
\int_{0}^{\infty} \d x\,  J^2_{m}(x\,v)\,
\mathrm{e}^{-\l[(x^2/2)+x\,\cEb\r]\,\bar{T}^2}\,
\mathrm{cosh}\l[m\,(x+\cEb)\,\bar{T}^2\r]\Biggr\},
\end{eqnarray}
\end{widetext}
where we have defined the dimensionless time interval~$\bar{T} =\gamma\,\Omega\,T$. 
The above integrals and sum seem difficult to evaluate analytically, but 
they can be computed numerically. 
We find that the functions $J_m^2(x\,v)$ that appear in the integrands 
behave as $x^{2\,m}$ when $x\to 0$, and as~$x^{-1}$ when~$x \to \infty$
(see, for instance, Refs.~\cite{watson2015treatise,BesselJ:large-order}).
Note that the factors~$\mathrm{e}^{-[(x^2/2)+x\,\cEb]\,\bar{T}^2}$ 
and~$\mathrm{cosh}\l[m\,(x+\cEb)\,\bar{T}^2\r]$ reduce to constants
as $x\to 0$.
As a result, the integrals are well behaved at small~$x$.
Moreover, at large~$x$, the integrals over~$x$ are dominated by the 
factor~$\mathrm{e}^{-x^2\,\bar{T}^2/2}$, and hence converge quickly
(in this regard, also see App.~\ref{app:convergence-int-x}).
We evaluate the integrals up to a suitably large value of $x$ and 
then carry out the sum involved.
In fact, we find that, because of the factor $\mathrm{e}^{-m^2\,\bar{T}^2/2}$,
the sum too converges extremely quickly. 
In Fig.~\ref{fig:RT11-fT-Mink-Vv}, we have plotted the transition probability rate
of the detector in circular motion for a given value of the velocity~$v$ and a few 
different values of the dimensionless time interval~$\bar{T}$.
Note that, when $\bar{T}$ is made larger, as expected, the transition probability
rate approaches the response rate of the detector that remains switched on forever.
Interestingly, for a given energy $\cEb$ and velocity~$v$, the transition 
probability rate of a detector that is switched on for a shorter duration 
is higher. 
However, this seems to occur up to a specific value of $\cEb$. 
At sufficiently large values of $\cEb$ and $\bar{T}$, the 
transition rate $R_{T}(\cEb)$ decreases, and all the curves corresponding to 
different values of $\bar{T}$ merge with the transition probability rate of
the detector that remains switched on forever.


\subsubsection{Response in a thermal bath}

Let us now evaluate the finite time response of the detector in circular motion
when it is immersed in a thermal bath.
As earlier, we shall consider Gaussian switching functions [cf. Eq.~\eqref{eq:gsf}].
On substituting the Wightman function at a finite temperature along the 
trajectory of the rotating detector [cf. Eq.~\eqref{eq:Greenfn-2UDD-thermal-i}] 
in the expression~\eqref{eq:tprT} that governs the transition probability rate 
of the detector, we obtain that
\begin{eqnarray}
R_T(\cE) 
&=& \int_{0}^{\infty} \f{\d q}{4\,\pi}\, 
\sum_{m=-\infty}^{\infty} J^2_{m}(q\,\sigma)\nn\\
& &\times\int_{-\infty}^{\infty}\d u\, 
\biggl\{\f{\mathrm{e}^{-i\,\l[\cE+\gamma\,(q-m\,\Omega)\r]\,u}\, 
\mathrm{e}^{- u^2/(2\,T^2)}}{1-\mathrm{e}^{-\beta\, q}}\nn\\
& & +\, \f{\mathrm{e}^{-i\,\l[\cE-\gamma\,(q-m\,\Omega)\r]\,u}\,
\mathrm{e}^{-u^2/(2\,T^2)}}{\mathrm{e}^{\beta\, q}-1}\biggr\}.
\end{eqnarray}
Upon carrying out the Gaussian integral over~$u$, we arrive at 
\begin{eqnarray}
R_T(\cE) 
&=&\sqrt{2\,\pi}\,T\, \int_{0}^{\infty} \f{\d q}{4\,\pi}\, 
\sum_{m=-\infty}^{\infty} J^2_{m}(q\,\sigma)\nn\\
& &\times\,\biggl\{\f{\mathrm{e}^{-\l[\cE+\gamma\,(q-m\,\Omega)\r]^2\, 
T^2/2}}{1-\mathrm{e}^{-\beta\, q}}\nn\\
& & +\, \f{\mathrm{e}^{-\l[\cE-\gamma\,(q-m\,\Omega)\r]^2\,
T^2/2}}{\mathrm{e}^{\beta\, q}-1}\biggr\}.
\end{eqnarray}
One can further simplify this expression to eventually obtain that
\begin{widetext}
\begin{eqnarray}\label{eq:response-fn-finite-thermal}
R_T(\cEb) &=& \f{\sqrt{2\,\pi}\,\bar{T}}{4\,\pi\gamma}\,
\mathrm{e}^{-\cEb^2\,\bar{T}^2/2}\,\Biggl\{\int_{0}^{\infty} \d x\,
J^2_{0}(x\,v)\,
\Biggl[\frac{\mathrm{e}^{-\l[(x^2/2)+x\,\cEb\r]\,\bar{T}^2}}{1
-\mathrm{e}^{-\bar{\beta}\,x}}
+ \f{\mathrm{e}^{-\l[(x^2/2)
-x\,\cEb\r]\,\bar{T}^2}}{\mathrm{e}^{\bar{\beta}\,x}-1}\Biggr]\nn\\
& &+\, 2\,\sum_{m=1}^{\infty}\,  \mathrm{e}^{-m^2\,\bar{T}^2/2}
\int_{0}^{\infty} \d x\, J^2_{m}(x\,v)\,
\Biggl[\f{\mathrm{e}^{-\l[(x^2/2)+x\,\cEb\r]\,\bar{T}^2}}{1
-\mathrm{e}^{-\bar{\beta}\,x}}\,
\mathrm{cosh}\l[m\,(x+\cEb)\,\bar{T}^2\r]\nn\\ 
& &+\, \f{\mathrm{e}^{-\l[(x^2/2)-x\,\cEb\r]\,\bar{T}^2}}{\mathrm{e}^{\bar{\beta}\,x}
-1}\,\mathrm{cosh}\l[m\,(x-\cEb)\,\bar{T}^2\r]\Biggr]\Biggr\}.
\end{eqnarray}
\end{widetext}

Recall that the functions $J_m^2(x\,v)$ behave as $x^{2\,m}$ when $x\to 0$.
Also, as we pointed out, the factors~$\mathrm{e}^{-[(x^2/2)+x\,\cEb]\,\bar{T}^2}$ 
and~$\mathrm{cosh}\l[m\,(x+\cEb)\,\bar{T}^2\r]$ reduce to constants when 
$x\to 0$.
Moreover, note that the functions $(1-\mathrm{e}^{-\bar{\beta}\,x})$ and 
$(\mathrm{e}^{\bar{\beta}\,x}-1)$ behave as $x$ when $x\to 0$. 
Therefore, in the $m=0$ term, as $x\to 0$, the integrand in the above 
expression for $R_T(\cEb)$ behaves as $x^{-1}$ and the integration over 
$x$ leads to a logarithmic divergence as $x\to 0$.
This is the infrared divergence at the finite temperature which we had 
discussed earlier.
In contrast to the case wherein the detectors are switched on forever,
it proves to be more involved to handle the divergence when the
detectors are switched on for a finite duration.
We shall regulate the divergence using a procedure which ensures that we 
recover the results we have already obtained in the limits $\bar{T}\to 
\infty$ and $\bar{\beta} \to \infty$.
In order to avoid a long digression, we have discussed the procedure in 
App.~\ref{app:remedy-infrared-div}.
Once we have tackled the divergence, the remaining terms can be evaluated 
without any difficulty.
As in the earlier cases, we have evaluated the integrals and sum in 
Eq.~\eqref{eq:response-fn-finite-thermal} numerically to arrive at 
the response of the detector. 
We had pointed out that, even in the Minkowski vacuum, 
the integrals at large~$x$ and the sum at large~$m$ are dominated by the 
factors~$\mathrm{e}^{-x^2\,\bar{T}^2/2}$ and~$\mathrm{e}^{-m^2\,\bar{T}^2/2}$,
respectively.
Hence, they converge very quickly, making it convenient to compute them 
numerically. 
In the finite temperature case of our interest, additionally, the contribution 
due to the exponential factor in the denominator in the final term leads to a 
more rapid convergence of the integral.
We have checked that the results are robust against increasing the upper limits 
of the integral over~$x$ and the sum over~$m$.

In Fig.~\ref{fig:RT11-fT-Mink-Vv}, we have plotted the transition probability
rate of the detector in circular motion and is immersed in a thermal bath for 
different values of the time interval~$\bar{T}$ for which the detector remains
effectively switched on, assuming a given velocity~$v$ and inverse 
temperature~$\beta$.
\begin{figure*}
\centering
\includegraphics[width=0.46\linewidth]{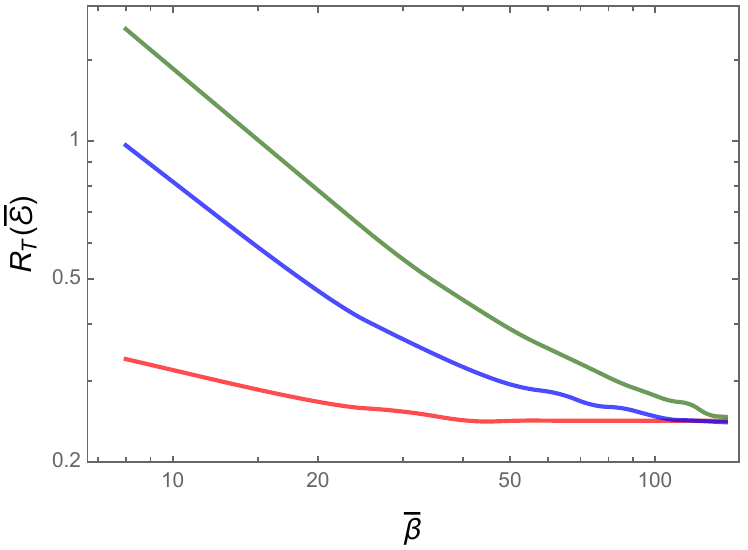}
\hskip 10pt
\includegraphics[width=0.475\linewidth]{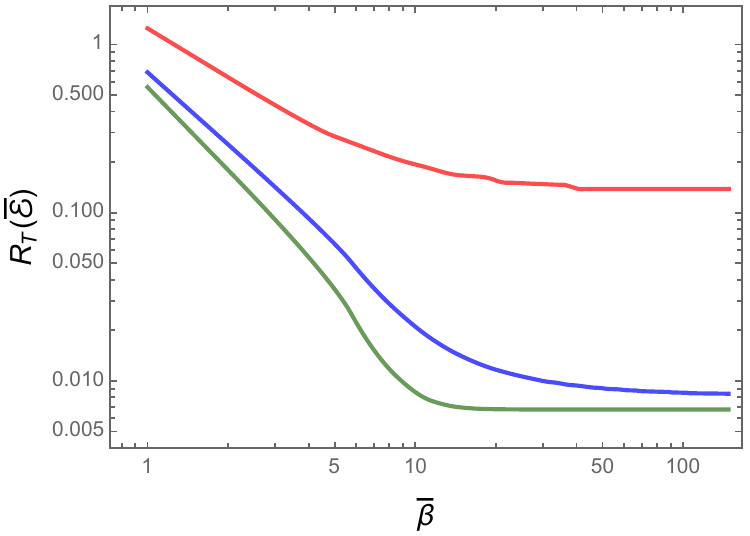}
\caption{The transition probability rate $R_{T}(\cEb,\bar{\beta})$ of the 
detector in circular motion that is immersed in a thermal bath and has been 
switched on for a finite time~$T$ has been plotted as a function
of the dimensionless inverse temperature~$\bar{\beta}$. 
We have set $v=0.5$, and have plotted the results for two different values of the 
dimensionless energy gap, viz. $\cEb= (0.02,0.6)$ (on the left and right), and three 
different values of switching time, viz. for $\bar{T}=(1, 5,10)$ (in red, blue and
green, respectively).
These plots clearly indicate that  the transition probability rate is larger 
for larger temperatures of the thermal bath (or, equivalently, for smaller
values of~$\bar{\beta}$).
We also observe that, when $\cEb$ is suitably large, the transition
probability rate increases with decreasing~$\bar{T}$, as in the case of the Minkowski
vacuum.
However, for smaller values of $\cEb$, we find that this behavior can be reversed 
when $\bar{\beta}$ is small.}\label{fig:R11-fT-Th-Vbeta}
\end{figure*}
In contrast to the response in the Minkowski vacuum, we find that, at a finite
temperature, the transition probability rate of the detector is lower at lower
energies, when the detector is switched on for a shorter duration.
It is only at high enough energies~$\cEb$ that the response is higher when~$\bar{T}$
is made smaller.
This point should also be clear from Fig.~\ref{fig:R11-fT-Th-Vbeta} wherein we 
have plotted the transition probability rate of the detector as a function of 
the dimensionless inverse temperature~$\bar{\beta}$ for different values of~$\cEb$ 
and $\bar{T}$.

We should point out here that, in the limit $\bar{\beta}\to\infty$, the quantities
$[1-\mathrm{exp}\,({-\bar{\beta}\,x})]^{-1}$ and $[\mathrm{exp}\,({\bar{\beta}\,x})
-1]^{-1}$ in Eq.~\eqref{eq:response-fn-finite-thermal} reduce to unity and zero,
respectively. 
Note that the limit $\bar{\beta}\to\infty$ implies a vanishing temperature for 
the thermal bath and hence corresponds to the Minkowski vacuum.
In such a case, the transition probability rates~\eqref{eq:response-fn-infinite-thermal} 
and~(\ref{eq:response-fn-finite-thermal}) reduce to the corresponding 
results for the Minkowski vacuum, viz. Eqs.~\eqref{eq:response-fn} 
and~\eqref{eq:response-fn-finite-Minkowski}, as required. 
Moreover, recall that, earlier, in Eq.~\eqref{eq:response-fn-infinite-thermal2}, we 
had expressed the transition probability rate of the rotating detector as a sum of 
different contributions arising due to spontaneous and stimulated excitations.
We should point out here that the transition probability rate of the rotating
detector which has been switched on for a finite time interval can also be
expressed in a similar fashion.


\section{Radiative processes of entangled detectors in circular 
motion}\label{sec:erd-mv-tb}

Having discussed the response of individual Unruh-DeWitt detectors,
let us now turn to examine the responses of two entangled detectors that
are moving on circular trajectories.


\subsection{Detectors switched on for infinite duration}\label{subsec:erd-mv-tb-IfT}

As we have done earlier, let us first discuss the case wherein the detectors
are switched on forever, before going to study the situations wherein the 
detectors are switched on for a finite time interval.


\subsubsection{Response in the Minkowski vacuum}

Consider two detectors that are moving along the circular trajectories 
$\tx_j=(\gamma_j\,\tau_j,\sigma_j,\gamma_j\,\Omega_j\,\tau_j)$
and $\tx_l=(\gamma_l\,\tau_l,\sigma_l,\gamma_l\,\Omega_l\,\tau_l)$,
with $v_{j(l)}=\sigma_{j(l)}\,\Omega_{j(l))}$ and $\gamma_{j(l)}
=1/\sqrt{1-v_{j(l)}^2}$.
The circular trajectories of the two detectors are, in general,
assumed to be independent.
On utilizing the mode 
decomposition~\eqref{eq:scalar-field-decomposition}
of the scalar field, the positive frequency Wightman function in the 
Minkowski vacuum which connects two spacetime points corresponding to two 
such detectors in circular motion can be obtained to be
\begin{eqnarray}\label{eq:Greenfn-2diff-jl}
G^{+}_{jl}\l[x_{j}(\tau_{j}),x_{l}(\tau'_{l})\r]
&=& \int_{0}^{\infty}\f{\d q}{4\,\pi}
\sum_{m=-\infty}^{\infty} J_{m}(q\,\sigma_{j})\, J_{m}(q\,\sigma_{l})\nn\\ 
& &\times\, \mathrm{e}^{-i\, q\, (\gamma_{j}\,\tau_{j}-\gamma_{l}\,\tau'_{l})}\nn\\
& &\times\,\mathrm{e}^{i\,m\,(\gamma_{j}\,\Omega_{j}\,\tau_{j}
-\gamma_{l}\,\Omega_{l}\,\tau'_{l})}.
\end{eqnarray}
We can arrive at the total transition probability~\eqref{eq:Transition-prob} 
corresponding to the entangled detectors by evaluating the auto and cross
transition probabilities~$F_{jl}(\cE)$ defined in Eq.~\eqref{eq:Transition-coeff}.

To evaluate the auto and cross transition probabilities, let us consider the change 
of variables $\bar{u}= \tau_{j} -\tau'_{l}$ and $\bar{v}=\tau_{j} + \tau'_{l}$.  
In terms of these new variables, for the case of detectors that are switched on
forever, i.e. when $\kappa_j(\tau_j)=\kappa_l(\tau_l)=1$, the transition probabilities 
$F_{jl}(\cE)$ can be expressed as
\begin{equation}
F_{jl}(\cE) 
= \int_{-\infty}^{\infty}\f{\d\bar{v}}{2}\, \int_{-\infty}^{\infty}\d\bar{u}\,
\mathrm{e}^{-i\,\cE\,\bar{u}}\, G^{+}_{jl}(\bar{u},\bar{v}).
\label{eq:Fjl-2UDD-1}
\end{equation}
Upon using the inverse transformations $\tau_{j} = (\bar{v} + \bar{u})/2$ and 
$\tau'_{l} = (\bar{v} - \bar{u})/2$ in the expression~\eqref{eq:Greenfn-2diff-jl},
we obtain the Wightman function $G^{+}_{jl}(\bar{u},\bar{v})$ for the two detectors 
moving on circular trajectories to be
\begin{eqnarray}\label{eq:Greenfn-ub-vb}
G^{+}_{jl}(\bar{u},\bar{v})
&=& \int_{0}^{\infty}\frac{\d q}{4\,\pi}\sum_{m=-\infty}^{\infty}
J_{m}(q\,\sigma_{j})\, J_{m}(q\,\sigma_{l})\nn\\ 
& &\times\, \mathrm{e}^{-i\, [\alpha_{1}(q)\,\bar{v} + \alpha_{2}(q)\,\bar{u}]/2},
\end{eqnarray}
where the quantities $\alpha_1(q)$ and $\alpha_2(q)$ are given by
\begin{subequations}
\label{eq:a1a2}
\begin{eqnarray}
\alpha_{1}(q) &=& q\,(\gamma_{j}-\gamma_{l}) - m\,(\gamma_{j}\, \Omega_{j}
-\gamma_{l}\, \Omega_{l}),\label{eq:a1}\\
\alpha_{2}(q) &=& q\,(\gamma_{j}+\gamma_{l}) - m\, (\gamma_{j}\, \Omega_{j}
+\gamma_{l}\, \Omega_{l}).
\end{eqnarray}
\end{subequations}
On carrying the integral over $\bar{u}$, we obtain the transition
probability~$F_{jl}(\cE)$ to be
\begin{eqnarray}\label{eq:Fjl-2UDD-2}
F_{jl}(\cE) &=& \int_{0}^{\infty}\f{\d q}{2} \sum_{m=-\infty}^{\infty}
J_{m}(q\,\sigma_{j})\, J_{m}(q\,\sigma_{l})\nn\\
& &\times\,\delta^{(1)}\l[\cE+\frac{\alpha_{2}(q)}{2}\r]\, 
\int_{-\infty}^{\infty}\f{\d\bar{v}}{2}\,
\mathrm{e}^{-i\,\alpha_{1}(q)\,\bar{v}/2}\nn\\
&=& \f{1}{\gamma_{j} + \gamma_{l}}\, \sum_{m\ge\hat{\cE}}^{\infty}
J_{m}(q_{0}\,\sigma_{j})\, J_{m}(q_{0}\,\sigma_{l})~\nn\\
& &\times\,\int_{-\infty}^{\infty}\frac{\d\bar{v}}{2}\,
\mathrm{e}^{-i\, \alpha_{1}(q_{0})\,\bar{v}/2},
\end{eqnarray}
where the quantity $q_0$ is defined as
\begin{equation}
q_{0} = \f{(m-\hat{\cE})}{\gamma_{j}+ \gamma_{l}}\,
(\gamma_{j}\,\Omega_{j} + \gamma_{l}\,\Omega_{l})\label{eq:q0}
\end{equation}
with $\hat{\cE}$ being given by
\begin{equation}
\hat{\cE} = \f{2\,\cE}{\gamma_{j}\,\Omega_{j} + \gamma_{l}\,\Omega_{l}}.
\label{eq:cEhat}
\end{equation}
We should point out that the condition $m\geq \hat{\cE}$
arises since~$q_0\geq 0$.
Let us now explicitly evaluate the transition probabilities $F_{jl}(\cE)$
for the cases $j=l$ and $j\neq l$.

When $j=l$, the trajectories correspond to the same detector so that we 
have $\gamma_{j} = \gamma_{l}$ and $\Omega_{j} = \Omega_{l}$, which lead 
to $\alpha_{1}(q_{0}) = 0$.
In such a case, the expression~\eqref{eq:Fjl-2UDD-2} reduces to the
transition probability of a single detector. 
We can also define the corresponding transition probability rate, say, 
$R_{jj}(\cE)$, by dividing the quantity $F_{jj}(\cE)$ in 
Eq.~\eqref{eq:Fjl-2UDD-2} by the integral over $\bar{v}$ [cf. Eq.~\eqref{eq:tpr}].
Since, $\hat{\cE} = \cE/(\gamma_{j}\,\Omega_{j}) = \bar{\cE}_{j}$ and
$q_{0}=(m-\bar{\cE}_{j})\,\Omega_{j}$ when $j=l$, we obtain the transition 
probability rate of the detector to be
\begin{equation}\label{eq:R11-2UDD}
R_{jj}(\cEb_j) = \f{1}{2\,\gamma_{j}}\, \sum_{m\ge\bar{\cE}_{j}}^{\infty}
J^2_{m}\l[(m-\cEb_{j})\,v_{j}\r],
\end{equation}
where, as we mentioned before, the condition $m\ge\bar{\cE}_{j}$ arises 
because $q_0\ge 0$.
This is exactly the result we had obtained earlier when we had considered 
the response of a single detector [cf. Eq.~\eqref{eq:response-fn}].
When $j=1$, the transition probability rate $R_{11}(\cEb_1)$ is given by 
the expression~\eqref{eq:response-fn}, with $(v, \gamma,\cEb$)
replaced by $(v_1,\gamma_{1},\cEb_{1})$. 
If we define the dimensionless parameters $\bar{\Omega} = \Omega_{2}/\Omega_{1}$
and $\bar{\gamma} = \gamma_{2}/\gamma_{1}$, then the transition probability
$R_{22}(\cEb_1)$ can be expressed as
\begin{equation}\label{eq:R11-2UDD}
R_{22}(\cEb_1) = \f{1}{2\,\gamma_{2}}\,
\sum_{m\ge\bar{\cE}_{1}/(\bar{\gamma}\,\bar{\Omega})}^{\infty}
J^2_{m}\l[\l(m-\f{\bar{\cE}_{1}}{\bar{\gamma}\,\bar{\Omega}}\r)\,v_2\r].
\end{equation}

Let us now consider the case wherein $j\neq l$. 
When $j\neq l$, the integral over $\bar{v}$ in Eq.~\eqref{eq:Fjl-2UDD-2}
results in a delta function of the form $\delta^{(1)}[\alpha_1(q_0)]$, which
can be utilized to define the transition probability rate.
Also, the delta function leads to an additional constraint on~$m$. 
For the transition probability rate $R_{jl}(\cE)$ to be non-zero, other than
the condition $m\ge\hat{\cE}$,  we also require that
\begin{equation}
m= m_{0} = \f{(\gamma_{l} - \gamma_{j})\,\cE}{\gamma_{j}\,
\gamma_{l}\,(\Omega_{j} - \Omega_{l})}.\label{eq:m0}
\end{equation}
In other words, the contribution to the transition probability rate 
arises due to only one term in the sum over~$m$, leading to
\begin{eqnarray}\label{eq:Rjl-2UDD}
R_{jl}(\hat{\cE}) = \f{1}{(\gamma_{j} + \gamma_{l})}\,
J_{m_{0}}(q_{0}\,\sigma_{j})\, J_{m_{0}}(q_{0}\,\sigma_{l}),
\end{eqnarray}
and we should stress that this result is true {\it only}\/ when $j\ne l$.
However, since $m_0$ has to be an integer, the relation~\eqref{eq:m0} implies 
that it is only for some specific values of the parameters of the system 
that the transition probability rates $R_{12}(\cE)$ and $R_{21}(\cE)$
contribute to the radiative process.
In terms of the dimensionless parameters $\bar{\gamma}$ and~$\bar{\Omega}$, 
we can express the transition probability rate $R_{12}(\cEb_1)$ as follows:
\begin{eqnarray}\label{eq:Rjl-2UDD-2}
R_{12}(\cEb_1) &=& \f{1}{(\gamma_{1} + \gamma_{2})}\,
J_{m_{0}}\l[(m_{0}-\hat{\cE})\,\frac{(1+\bar{\gamma}\, 
\bar{\Omega})}{(1+\bar{\gamma})}\,v_1\r]\nn\\
& &\times\,J_{m_{0}}\l[(m_{0}-\hat{\cE})\,
\f{(1+\bar{\gamma}\,\bar{\Omega})}{(1+\bar{\gamma})\,\bar{\Omega}}\,v_2\r]\nn\\
&=& R_{21}(\cEb_1)
\end{eqnarray}
with $\hat{\cE}$ and $m_0$ being given by
\begin{eqnarray}\label{eq:m0-expression}
\hat{\cE} &=& \f{2\,\cE}{\gamma_{1}\,\Omega_{1} + \gamma_{2}\,\Omega_{2}} 
= \f{2\,\cEb_{1}}{1 + \bar{\gamma}\,\bar{\Omega}},\nn\\
m_{0} &=& \f{(\gamma_{2} - \gamma_{1})\,\cE}{\gamma_{1}\,\gamma_{2}\,(\Omega_{1}\,
- \Omega_{2})} = \frac{(1 - 1/\bar{\gamma})\,\cEb_{1}}{1 - \bar{\Omega}}.
\end{eqnarray}

Let us now understand if the transition probability rate $R_{12}(\cEb_1)$ can 
be non-zero for values of the parameters describing the trajectories of the 
two detectors, viz. $(\Omega_1,\gamma_1)$ and $(\Omega_2,\gamma_2)$, that 
we shall focus on.
First, consider the case wherein $\gamma_1\ne \gamma_2$, while $(\Omega_1 - 
\Omega_2)\to 0^{+}$.
In such a situation, $m_0 \to \infty$ as $\Omega_1\to\Omega_2$, 
and the required condition $m_0\ge \hat{\cE}$ will indeed be satisfied. 
However, we find that, as $m\to \infty$, the function $J_m(z)$ goes to 
zero (in this context, see Refs.~\cite{watson2015treatise,BesselJ:large-order}).
This implies that the transition probability rate $R_{12}(\cE)$ vanishes.
Second, when $\gamma_1=\gamma_2$, we require $\Omega_1\ne \Omega_2$ so that they
correspond to different trajectories for the two detectors.
In such a situation, $m_0=0$.
However, since $\hat{\cE}>0$, the condition $m_0\ge \hat{\cE}$ cannot be 
satisfied leading to a vanishing $R_{12}(\cEb_1)$. 
Hence, in these situations, the complete transition probability rate of the 
two entangled detectors will be solely determined by the rates $R_{11}(\cEb_1)$ 
and $R_{22}(\cEb_1)$ of the individual detectors.
Naturally, constructive or destructive effects due to the cross transition
probability rates $R_{12}(\cEb_1)$ and $R_{21}(\cEb_1)$ will be absent in 
the corresponding total transition rate.
However, in general, non-zero contributions due to $R_{12}(\cEb_1)$ 
and $R_{21}(\cEb_1)$ can be expected to arise when one considers, say, 
transitions from the symmetric and anti-symmetric Bell states to the 
collective excited state.
We shall discuss these points further in the concluding section.


\subsubsection{Response in a thermal bath}\label{sec:red-tb}

Let us now evaluate the response of the entangled detectors in a thermal
bath.
In a thermal bath, upon using the 
decomposition~\eqref{eq:scalar-field-decomposition} of the scalar field, 
one can obtain the positive frequency Wightman function connecting two 
spacetime points corresponding to two differently rotating detectors, 
denoted by the subscripts~$j$ and~$l$, to be
\begin{widetext}
\begin{eqnarray}\label{eq:Greenfn-2UDD-thermal}
G^{+}_{\beta_{jl}}\l[\tx_{j}(\tau_{j}),\tx_{l}(\tau'_{l})\r] 
&=& \int_{0}^{\infty}\f{\d q}{4\,\pi}\, \sum_{m=-\infty}^{\infty}
J_{m}(q\,\sigma_{j})\, J_{m}(q\,\sigma_{l})\nn\\
& &\times\, \biggl[\f{\mathrm{e}^{-i\, [q\, (\gamma_{j}\,\tau_{j}
-\gamma_{l}\,\tau'_{l})
- m\,(\gamma_{j}\,\Omega_{j}\,\tau_{j}-\gamma_{l}\,\Omega_{l}\,
\tau'_{l})]}}{1-\mathrm{e}^{-\beta\, q}}
+ \f{\mathrm{e}^{i\,[q\, (\gamma_{j}\,\tau_{j}-\gamma_{l}\,\tau'_{l})
-m\,(\gamma_{j}\,\Omega_{j}\,\tau_{j}-\gamma_{l}\,\Omega_{l}\,
\tau'_{l})]}}{\mathrm{e}^{\beta\,q}-1}\biggr].
\end{eqnarray}
To calculate the corresponding transition probabilities~$F_{jl}(\cE)$ 
[cf. Eq.~\eqref{eq:Transition-coeff}], we proceed as in the case of 
the Minkowski vacuum and consider the change of variables $\bar{u}
= \tau_{j} -\tau'_{l}$ and $\bar{v}=\tau_{j} + \tau'_{l}$. 
With the change of variables, the above Wightman function simplifies 
to be
\begin{eqnarray}\label{eq:W-fn-tb-ub-vb}
G^{+}_{\beta_{jl}}(\bar{u},\bar{v})
= \int_{0}^{\infty}\frac{\d q}{4\,\pi}\, \sum_{m=-\infty}^{\infty}
J_{m}(q\,\sigma_{j})\, J_{m}(q\,\sigma_{l})\,
\l[\f{\mathrm{e}^{-i\, \l[\alpha_{1}(q)\,\bar{v} 
+ \alpha_{2}(q)\,\bar{u}\r]/2}}{1-\mathrm{e}^{-\beta\, q}}
+\f{\mathrm{e}^{i\,\l[\alpha_{1}(q)\,\bar{v} 
+ \alpha_{2}(q)\,\bar{u}\r]/2}}{\mathrm{e}^{\beta\,q}-1}\r],
\end{eqnarray}
where the quantities $\alpha_1(q)$ and $\alpha_2(q)$ are given by 
Eq.~\eqref{eq:a1a2}.
Upon substituting this expression in Eq.~\eqref{eq:Transition-coeff}
and integrating over $\bar{u}$, we obtain the transition 
probabilities~$F_{jl}(\cE)$ to be
\begin{eqnarray}\label{eq:Fjl-2UDD-T1}
F_{jl}(\cE) &=& \int_{0}^{\infty}\f{\d q}{2}\, \sum_{m=-\infty}^{\infty}
J_{m}(q\,\sigma_{j})\, J_{m}(q\,\sigma_{l})\,
\biggl\{\f{\delta^{(1)}\l[\cE+\alpha_{2}(q)/2\r]}{1-\mathrm{e}^{-\beta\, q}} 
+ \f{\delta^{(1)}\l[-\cE+\alpha_{2}(q)/2\r]}{\mathrm{e}^{\beta\, q}-1}\biggr\}\,
\int_{-\infty}^{\infty}\f{\d\bar{v}}{2}\,
\mathrm{e}^{-i\, \alpha_{1}(q)\,\bar{v}/2}\nn\\
&=& \f{1}{(\gamma_{j} + \gamma_{l})}\, 
\biggl[\sum_{m\ge\hat{\cE}}^{\infty}
\f{J_{m}(q_{0}\,\sigma_{j})\, J_{m}(q_{0}\,\sigma_{l})}{1
-\mathrm{e}^{-\beta\, q_{0}}}
\int_{-\infty}^{\infty}\f{\d\bar{v}}{2}\,
\mathrm{e}^{-i\, \alpha_{1}(q_{0})\,\bar{v}/2}\nn\\
& &+\, \sum_{m\ge -\hat{\cE}}^{\infty}
\f{J_{m}(\bar{q}_{0}\,\sigma_{j})\, J_{m}(\bar{q}_{0}\,
\sigma_{l})}{\mathrm{e}^{\beta\,\bar{q}_{0}}-1}\,
\int_{-\infty}^{\infty}\frac{\d\bar{v}}{2}\,
\mathrm{e}^{-i\, \alpha_{1}(\bar{q}_{0})\,\bar{v}/2}\biggr],
\end{eqnarray}
\end{widetext}
where $q_0$ is given by Eq.~\eqref{eq:q0}, while $\bar{q}_0$ is defined to 
be
\begin{equation}
\bar{q}_{0} = \f{(m+\hat{\cE})}{\gamma_{j} + \gamma_{l}}\,
(\gamma_{j}\,\Omega_{j} + \gamma_{l}\,\Omega_{l})
\end{equation}
and, as before, $\hat{\cE}$ is given by Eq.~\eqref{eq:cEhat}.

Let us first discuss the results in the cases wherein $j=l$ and $j\ne l$.
When $j=l$, we have $\alpha_{1} (q_0)= \alpha_{1} (\bar{q}_0)=0$ [cf. 
Eq.~\eqref{eq:a1}], and one can readily determine the corresponding 
transition probability rate to be
\begin{eqnarray}\label{eq:Rjj-2UDD-T1}
R_{jj}(\cEb_j) 
&=& \f{1}{2\,\gamma_{j}}\, \biggl\{\sum_{m\ge\bar{\cE}_{j}}^{\infty}
\f{J^2_{m}\l[(m-\cEb_{j})\,v_{j}\r]}{1-\mathrm{e}^{-\bar{\beta}_{j}\,
(m-\cEb_{j})}}\nn\\
& &+\,\sum_{m\ge -\bar{\cE}_{j}}^{\infty}
\frac{J^2_{m}\l[(m+\cEb_{j})\,v_{j}\r]}{\mathrm{e}^{\bar{\beta}_{j}\, 
(m+\cEb_{j})}-1}\biggr\},
\end{eqnarray}
where we have set $\bar{\beta}_{j} =\beta\,\Omega_{j}$.
As expected, this result is same as the transition probability rate of 
a single detector in a thermal bath that we had obtained earlier.
For instance, the transition probability rate $R_{11}(\cEb_1)$ is given
by the expression~\eqref{eq:response-fn-infinite-thermal}, with $(v, 
\gamma,\cEb,\bar{\beta})$ replaced by $(v_1,\gamma_{1},\cEb_{1},\bar{\beta}_1)$.
Also, we can express the transition probability rate $R_{22}(\cEb_1)$ 
in terms of the dimensionless parameters $\bar{\Omega}$ and $\bar{\gamma}$
as follows:
\begin{eqnarray}
R_{22}(\cEb_1) 
&=& \f{1}{2\,\gamma_2}\, 
\biggl\{\sum_{m\ge\bar{\cE}_{1}/(\bar{\gamma}\,\bar{\Omega})}^{\infty}
J^2_{m}\l[\l(m-\f{\bar{\cE}_{1}}{\bar{\gamma}\,\bar{\Omega}}\r)\,v_{2}\r]\nn\\
& &\times\,\f{1}{1-\mathrm{e}^{-\bar{\beta}_{1}\,\bar{\Omega}\,
[m-(\bar{\cE}_{1}/\bar{\gamma}\,\bar{\Omega})]}}\nn\\
& &+\,\sum_{m\ge -\bar{\cE}_{1}/(\bar{\gamma}\,\bar{\Omega})}^{\infty}
J^2_{m}\l[\l(m+\f{\bar{\cE}_{1}}{\bar{\gamma}\,\bar{\Omega}}\r)\,v_2\r]\nn\\
& &\times\,\f{1}{\mathrm{e}^{\bar{\beta}_{1}\,\bar{\Omega}\,
[m+(\bar{\cE}_{1}/\bar{\gamma}\,\bar{\Omega})]}-1}\biggr\}.
\end{eqnarray}

Let us now turn to the case wherein $j\neq l$.
In this case, we can notice from Eq.~\eqref{eq:Fjl-2UDD-T1} that 
the integral over $\bar{v}$ leads to $\delta^{(1)}[\alpha_1(q_0)]$
and $\delta^{(1)}[\alpha_1(\bar{q}_0)]$ in the first and the second 
sums, respectively.
This implies that non-trivial contributions arise only when $\alpha_{1}(q_{0})$ 
and $\alpha_{1}(\bar{q}_{0})$ vanish in these sums.
We find that, $\alpha_{1}(q_{0})=0$ corresponds to $m=m_{0}$ as have
encountered earlier in the Minkowski vacuum, whereas $\alpha_{1}(\bar{q}_{0}) 
= 0$ corresponds to $m=-m_{0}$. 
Note that, in order to lead to non-zero contributions to the transition 
probability rate~$R_{jl}(\cE)$ (with $j\neq l$), while the constraint $m_{0}
\ge \hat{\cE}$ must be fulfilled in the first sum, the condition $-m_{0}\ge
-\hat{\cE}$ must be satisfied in the second sum. 
It is easy to observe that these two conditions cannot be satisfied simultaneously 
as the second condition, which corresponds to $m_{0}\le \hat{\cE}$, is mathematically 
the opposite of the first one.

In particular, if the first condition is fulfilled, i.e. when $m_{0}\ge \hat{\cE}$, 
then the transition probability rate is given by
\begin{eqnarray}\label{eq:Rjl-2UDD-T1}
R_{12}(\cEb_1) &=& \f{1}{(\gamma_{1} + \gamma_{2})}\,
J_{m_{0}}\l[(m_{0}-\hat{\cE})\,
\f{(1+\bar{\gamma}\,\bar{\Omega})}{(1+\bar{\gamma})}\,v_1\r]\nn\\
& &\times\,J_{m_{0}}\l[(m_{0}-\hat{\cE})\,
\f{(1+\bar{\gamma}\,\bar{\Omega})}{(1+\bar{\gamma})\,\bar{\Omega}}\,v_2\r]\nn\\
& &\times\,\f{1}{1-\mathrm{e}^{-\bar{\beta}_{1}\,(m_{0}-\hat{\cE})\,
(1+\bar{\gamma}\,\bar{\Omega})/(1+\bar{\gamma})}}\nn\\
&=& R_{21}(\cEb_1),
\end{eqnarray}
where $\bar{\beta}_{1} = \beta\,\Omega_{1}$. 
As we had discussed before, we have $m_0=0$, when $\gamma_1=\gamma_2$ and 
$\Omega_1\ne \Omega_2$.
In such a case, the condition $m_0\ge \hat{\cE}$ will not be satisfied and 
hence $R_{12}(\cEb_1)$ vanishes. 
Moreover, when $\gamma_1\ne \gamma_2$ and $(\Omega_1 - \Omega_2)\to 0^{+}$, 
$m_0 \to \infty$.
Since $\hat{\cE}$ is finite, the condition $m_0\ge \hat{\cE}$ will indeed be satisfied. 
However, as we had pointed out earlier, when $m \to \infty$, the Bessel functions
$J_m(z)$ go to zero~\cite{watson2015treatise,BesselJ:large-order}.
As a result, the transition probability rate $R_{12}(\cE)$ vanishes in this case
as well.

\begin{figure}[!t]
\centering
\includegraphics[width=1\linewidth]{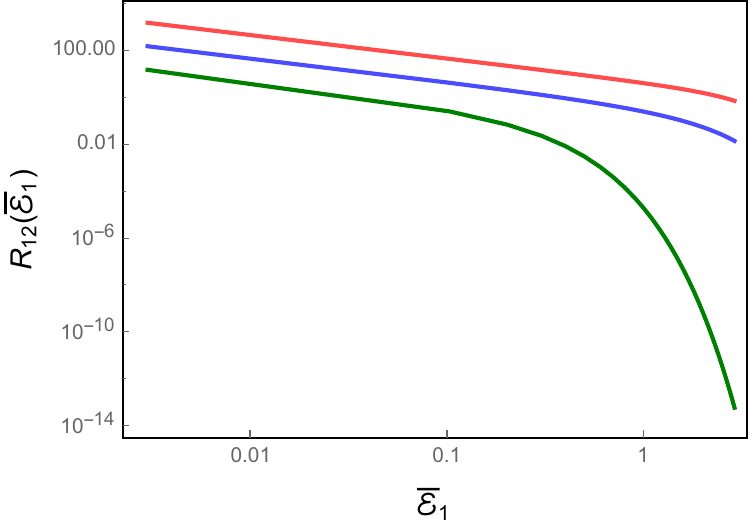}
\caption{The cross transition probability rate $R_{12}(\cEb_{1})$
of the two entangled detectors moving on circular trajectories and immersed in 
a thermal bath has been plotted as a function of~$\cEb_{1}$, for the case wherein
the detectors remain switched on forever.
We have set $v_1=v_2=0.5$ so that $\bar{\gamma}=1$, which corresponds to $m_0=0$,
and have chosen $\bar{\Omega}=5$. 
We have plotted the results for the cases wherein $\bar{\beta}_{1}=(0.1,1,10)$
(in red, blue, and green, respectively).
Clearly, the cross  transition probability rate is higher at a higher
temperature.}\label{fig:R12-InfT-Th-VEb1}
\end{figure}

On the other hand, when the second condition $m_{0}\le \hat{\cE}$ is satisfied,
the transition probability rate is given by
\begin{eqnarray}\label{eq:Rjl-2UDD-T2}
R_{12}(\cEb_1) &=& \f{1}{(\gamma_{1} + \gamma_{2})}\,
J_{m_{0}}\l[(-m_{0}+\hat{\cE})\,
\f{(1+\bar{\gamma}\,\bar{\Omega})}{(1+\bar{\gamma})}\,v_1\r]\nn\\
& &\times\,J_{m_{0}}\l[(-m_{0}+\hat{\cE})\,
\f{(1+\bar{\gamma}\,\bar{\Omega})}{(1+\bar{\gamma})\,\bar{\Omega}}\,v_2\r]\nn\\
& &\times\,\f{1}{\mathrm{e}^{\bar{\beta}_{1}\,(-m_{0}+\hat{\cE})\,
(1+\bar{\gamma}\,\bar{\Omega})/(1+\bar{\gamma})}-1}\nn\\
&=& R_{21}(\cEb_1).
\end{eqnarray}
Recall that, when $\gamma_1\neq \gamma_2$ and $\Omega_1\to \Omega_2$, $m_0\to
\infty$.
In such a case, clearly, the condition $m_{0}\le \hat{\cE}$ cannot be met and 
the cross transition probability rate $R_{12}(\cEb_1)$ between the two entangled
detectors will be zero.
But, for the case wherein $\gamma_1=\gamma_2$ and $\Omega_1\neq \Omega_2$,
since $m_0=0$, clearly, the condition $m_{0}\le \hat{\cE}$ 
is satisfied and the cross transition probability rate will be given by the 
above expression for~$R_{12}(\cEb_1)$.
Evidently, in contrast to the response in the Minkowski vacuum wherein the cross
transition probability rates were always zero (for the parameters
we focus on), these rates can contribute non-trivially in a thermal bath for 
certain sets of the parameters involved. 
In Fig.~\ref{fig:R12-InfT-Th-VEb1}, we have plotted the cross transition 
probability rate $R_{12}(\cEb_1)$ for a set of parameters that satisfy 
the condition $m_{0}\le \hat{\cE}$ and also correspond to an integer 
value for~$m_0$ (in fact, for $m_0=0)$.
We have plotted the rate for a few different values of the dimensionless inverse 
temperature~$\bar{\beta}_1$.
The figure suggests that the cross transition probability rate decreases with 
increasing $\bar{\beta}_{1}$, a behavior we had encountered earlier when we
had discussed the results for the case of a single detector, which corresponds
to the auto transition probability rate.
In the next section, we will discuss the complete transition probability
of the rotating detector, including the auto and the cross transition
probability rates.
As we shall see, the non-zero cross transition probability rates contribute  
constructively and destructively for the transition from the symmetric 
and anti-symmetric Bell states to the collective excited state. 

Lastly, it should be noted that, in the limit $\beta \to \infty$, i.e.
when the temperature of the thermal bath vanishes, the transition 
probability rate~$R_{12}(\cEb_1)$ as given by Eq.~\eqref{eq:Rjl-2UDD-T2} 
above reduces to zero.
In the same limit, the last factor in the transition probability 
rate~$R_{12}(\cEb_1)$ as given by Eq.~\eqref{eq:Rjl-2UDD-T1} simplifies 
to unity and the expression reduces to the result in the Minkowski 
vacuum [cf. Eq.~\eqref{eq:Rjl-2UDD-2}], as required.


\subsection{Detectors switched on for a finite duration}

Let us now turn to discuss the responses of entangled detectors that have been
switched on for a finite time interval.


\subsubsection{Response in the Minkowski vacuum}

As we have done earlier in the case of single detectors, let us now introduce 
Gaussian switching functions [cf. Eq.~\eqref{eq:gsf}] to examine the response
of entangled detectors that are switched on for a finite time interval.
In such a case, the transition probabilities~\eqref{eq:Transition-coeff} for 
two entangled detectors is given by
\begin{eqnarray}\label{eq:Fjl-FTM1}
F_{jl}^{T}(\mathcal{E}) 
&=& \int_{-\infty}^{\infty}\d\tau'_{l} \int_{-\infty}^{\infty}\d\tau_{j}\, 
\mathrm{e}^{-i\,\cE\,(\tau_{j}-\tau'_{l})}\, 
G^{+}\l(\tau_{j},\tau'_{l}\r)\nn\\
& &\times\, \exp \l[-(\tau^2_{j} + \tau'^2_{l})/T^2\r]
\end{eqnarray}
which, in terms of the variables $\bar{u}=(\tau_{j}-\tau'_{l})$ and $\bar{v}
=(\tau_{j}+\tau'_{l})$, can be expressed as
\begin{eqnarray}\label{eq:Fjl-FTM2}
F_{jl}^{T}(\mathcal{E}) &=& \int_{-\infty}^{\infty}\frac{\d\bar{v}}{2}\,
\mathrm{e}^{-\bar{v}^2/(2\,T^2)}\nn\\
& &\times\,\int_{-\infty}^{\infty}\d\bar{u}\,  
\mathrm{e}^{-i\,\mathcal{E}\,\bar{u}}\, 
G^{+}\l(\bar{u},\bar{v}\r)\,\mathrm{e}^{-\bar{u}^2/(2\,T^2)}.\qquad
\end{eqnarray}

Recall that, in the Minkowski vacuum, the Wightman function associated with the
two entangled detectors that are in circular motion can be expressed as in
Eq.~\eqref{eq:Greenfn-ub-vb}.
Also, as in the case of the single detector, we can define the transition probability 
rate of the detectors to be
\begin{equation}
R_{jl}^{T}(\mathcal{E}) = \f{F_{jl}^{T}(\mathcal{E})}{\sqrt{(\pi/2)}\,T}.     
\label{eq:tpr-ft}
\end{equation}
Upon substituting the Wightman function~\eqref{eq:Greenfn-ub-vb}  
in Eq.~\eqref{eq:Fjl-FTM2}, we find that the corresponding transition 
probability rate can be expressed as
\begin{eqnarray}\label{eq:Rjl-FTM1}
R_{jl}^{T}(\mathcal{E}) 
&=& \f{1}{\sqrt{(\pi/2)}\, T} 
\int_{0}^{\infty}\frac{\d q}{4\,\pi}
\sum_{m=-\infty}^{\infty} J_{m}(q\, \sigma_{j})\,J_{m}(q\,\sigma_{l})\nn\\
& &\times\,\int_{-\infty}^{\infty}\f{\d \bar{v}}{2}\, 
\mathrm{e}^{-\l[(\bar{v}^2/T^2)+ i\,\alpha_{1}(q)\, \bar{v}\r]/2}\nn\\
& &\times\, \int_{-\infty}^{\infty}\d \bar{u}\, \mathrm{e}^{-i\,\cE\,\bar{u}}\,
\mathrm{e}^{-\l[(\bar{u}^2/T^2)+ i\,\alpha_{2}(q)\, \bar{u}\r]/2}.
\end{eqnarray}
After carrying out the Gaussian integrals, we obtain that
\begin{eqnarray}\label{eq:Rjl-FTM2}
R_{jl}^{T}(\mathcal{E}) 
&=& \sqrt{2\,\pi}\,T \int_{0}^{\infty}\f{\d q}{4\,\pi}\,
\sum_{m=-\infty}^{\infty} J_{m}(q\, \sigma_{j})\,J_{m}(q\,\sigma_{l})\nn\\
& &\times\, \mathrm{e}^{-\l\{\alpha_{1}^2(q) + \l[\alpha_{2}(q)+2\,\mathcal{E}\r]^2\r\}\,T^2/8}.
\end{eqnarray}
Since the integral over $q$ and the sum over $m$ do not seem to be analytically
tractable, we need to compute them numerically as in the case of the single
detector.

Let us first consider the auto transition probability rates of the two detectors.
As we had discussed, when $j=l$, we have $\alpha_{1}(q) = 0$ and $\alpha_{2}(q) =
2\,\gamma_{j}(q-m\, \Omega_{j})$ and, in such a situation, the above 
transition probability rate reduces to
\begin{eqnarray}\label{eq:Rjj-FTM-S1}
R_{jj}^{T}(\mathcal{E}) 
&=& \sqrt{2\,\pi}\,T\,
\int_{0}^{\infty}\f{\d q}{4\,\pi} \sum_{m=-\infty}^{\infty}\, J^2_{m}(q\, \sigma_{j})\nn\\
& &\times\, \mathrm{e}^{-\l[\cE+\gamma_{j}\,(q-m\,\Omega_{j})\r]^2\,T^2/2}.
\end{eqnarray}
Note that, as expected, this result exactly matches the transition probability 
rate of a single detector [cf. Eq.~\eqref{eq:RT-single-detector}].
We can also introduce the dimensionless variable $x_{j}=q/\Omega_{j}$ and the
dimensionless parameter $\bar{T}_{j} = \gamma_j\,\Omega_{j}\,T$ to rewrite the
above integral and sum, as we have done earlier.
If we do so, we find that the result for $R_{11}^{T}(\cEb_1)$ is given by the 
expression in Eq.~\eqref{eq:response-fn-finite-Minkowski} with $(v,\gamma, 
\cEb, x, \bar{T})$ replaced by $(v_1,\gamma_1,\cEb_1, x_1, \bar{T}_1)$. 
With respect to the same set of dimensionless quantities as well as the 
dimensionless parameters $\bar{\gamma}$ and $\bar{\Omega}$, we find that 
the auto transition probability rate~$R_{22}^{T}(\cEb_1)$ can be expressed
as follows:
\begin{eqnarray}\label{eq:R22-FTM-S1}
R_{22}^{T}(\cEb_1) 
&=& \f{\sqrt{2\,\pi}\;\bar{T}_{1}}{4\,\pi\,\gamma_{1}}\,
\int_{0}^{\infty} \d x_{1}\,\sum_{m=-\infty}^{\infty}\,
J^2_{m}(x_{1}\,v_{2}/\bar{\Omega})\nn\\
& & \times\,\mathrm{e}^{-\l[\cEb_{1}+\bar{\gamma}\,
(x_1-\,m\,\bar{\Omega})\r]^2\,\bar{T}_1^2/2}.
\end{eqnarray}
We should point out that, when $\bar{\gamma}=1=\bar{\Omega}$ and 
$v_{2} = v_{1}$, this expression reduces to $R_{11}^{T}(\cEb_1)$. 

\begin{figure}[t]
\centering
\includegraphics[width=1\linewidth]{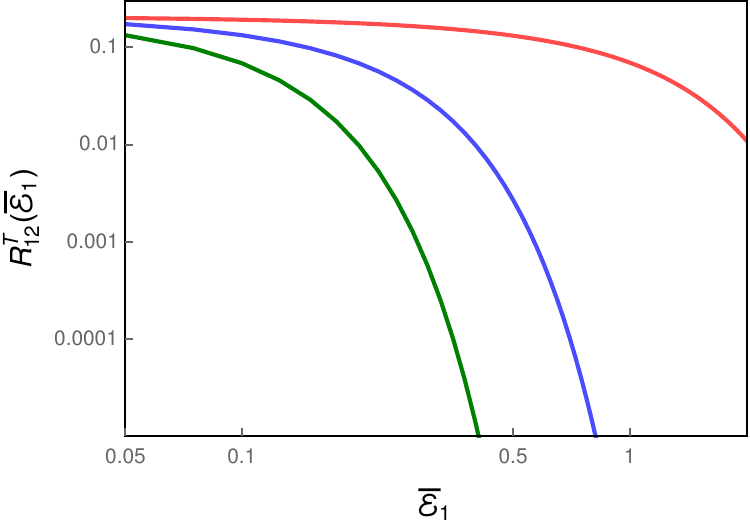}
\caption{The cross transition probability rate $R_{12}^T(\cEb_{1})$ 
of the two entangled detectors in the Minkowski vacuum has been plotted 
as a function of $\cEb_{1}$ when the detectors are in motion on circular 
trajectories and are switched on for a finite time interval~$T$.
As in the previous figure, we have set $v_1=v_2=0.5$ so that $\bar{\gamma}=1$,
and have chosen $\bar{\Omega}=5$. 
We have plotted the results for three different values of the dimensionless
time interval, viz. $\bar{T}_{1}=(1,5,10)$ (in red, blue, and green, 
respectively).
Note that, as in the case of the auto transition probability rate of a single 
detector (see Fig.~\ref{fig:RT11-fT-Mink-Vv}), the cross transition probability
rate $R_{12}(\cEb_{1})$ is higher when the detector is switched on for a 
shorter time interval.} \label{fig:R12-FiniteT-Mink-VEb1}
\end{figure}

When $j=1$ and $l=2$, we can express the cross transition probability 
rate $R_{12}^{T}(\cEb_1)$ as
\begin{eqnarray}\label{eq:R12-FTM-S1}
R_{12}^{T}(\cEb_1) 
\!\!&=&\!\! \f{\sqrt{2\,\pi}\, \bar{T}_{1}}{4\,\pi\,\gamma_{1}}\! 
\sum_{m=-\infty}^{\infty} \int_{0}^{\infty}\d x_{1}\, J_{m}(x_{1}\, v_{1})\,
J_{m}(x_{1} v_{2}/\bar{\Omega})\nn\\
& &\times\, \exp\,\biggl(-\biggl\{\l[(1-\bar{\gamma})\,x_1
-m\,(1-\bar{\gamma}\,\bar{\Omega})\r]^2\nn\\
& & + \l[2\,\cEb_{1}+(1+\bar{\gamma})\,x_1
-m\,(1+\bar{\gamma}\,\bar{\Omega})\r]^2\biggr\}\,\bar{T}_1^2/8\biggr)\nn\\
&=& R_{21}^{T}(\cEb_1).
\end{eqnarray}
This expression too reduces to that of $R_{11}^{T}(\mathcal{E})$ 
when $\bar{\gamma} = 1 = \bar{\Omega}$ and $v_{2} = v_{1}$.
In App.~\ref{app:tpr-ft-fe}, for convenience, we have explicitly listed the 
complete expressions for the finite time auto and cross transition probability 
rates $R_{11}^{T}(\cEb_1)$, $R_{22}^{T}(\cEb_1)$ and $R_{12}^{T}(\cEb_1)$.
It is these expressions that we actually utilize to numerically compute the 
rates.
While computing the rates, we work with the limits for the variables $x$ 
and $m$ for the integral and the sum that we had considered in the case 
of a single detector (in this context, see the caption 
of Fig.~\ref{fig:RT11-fT-Mink-Vv}).
In Fig.~\ref{fig:R12-FiniteT-Mink-VEb1} we have plotted the above cross
transition probability rate $R_{12}(\cEb_{1})$ for a given set of 
parameters describing the circular motion of the detector and for 
different values of the dimensionless time parameter $\bar{T}_1$.
We observe that the results for the cross transition probability rate are 
broadly similar to the auto transition probability rate we had obtained 
earlier in the case of a single detector (see Fig.~\ref{fig:RT11-fT-Mink-Vv}).
These plots also suggest that the transition probability rate is higher when 
the detectors interact with the field for a smaller time interval.


\subsubsection{Response in a thermal bath}\label{subsubsec:radiative-TB}

We can repeat the procedure that we have adopted earlier to determine the response
of two entangled detectors in circular motion and are immersed in a thermal bath.
For Gaussian switching functions, we can make use of the 
expression~\eqref{eq:Fjl-FTM2} for the transition probability of the detectors, with 
the Wightman function in the thermal bath being given by Eq.~\eqref{eq:W-fn-tb-ub-vb}.
Upon carrying out the resulting Gaussian integrals over $\bar{u}$ and $\bar{v}$, 
we find that the transition probability rate of the detectors [as defined in 
Eq.~\eqref{eq:tpr-ft}] can be expressed as
\begin{eqnarray}\label{eq:Rjl-FT-T2}
R_{jl}^{T}(\mathcal{E}) 
&=& \sqrt{2\,\pi}\,T 
\int_{0}^{\infty}\f{\d q}{4\,\pi}\, \sum_{m=-\infty}^{\infty}\,
J_{m}(q\, \sigma_{j})\,J_{m}(q\, \sigma_{l})\nn\\
& &\times\, \biggl\{\f{\mathrm{e}^{-\l\{\alpha_{1}^2(q)
+ [\alpha_{2}(q)+2\,\mathcal{E}]^2\r\}\,T^2/8}}{1-\mathrm{e}^{-\beta\, q}}\nn\\
& & +\,\f{\mathrm{e}^{-\l\{\alpha_{1}^2(q) 
+ [\alpha_{2}(q)-2\,\mathcal{E}]^2\r\}\,T^2/8}}{\mathrm{e}^{\beta\,q}-1}\biggr\}.
\end{eqnarray}

When $j=l$, since $\alpha_{1}(q) = 0$ and $\alpha_{2}(q) = 2\,\gamma_j\,(q-m\,
\Omega_j)$, the above transition probability rate can be expressed as
\begin{eqnarray}\label{eq:Rjj-FT-TB}
R_{jj}^{T}(\mathcal{E}) 
&=& \sqrt{2\,\pi}\,T\, \int_{0}^{\infty}\f{\d q}{4\,\pi}\,
\sum_{m=-\infty}^{\infty} J^2_{m}(q\, \sigma_{j})\nn\\
& &\times\, \bigg\{\frac{\mathrm{e}^{-\l[\cE+\gamma_j\,(q-m\,\Omega_j)\r]^2\,
T^2/2}}{1-\mathrm{e}^{-\beta\, q}}\nn\\
& &+\, \f{\mathrm{e}^{-\l[\cE-\gamma_j\,(q-m\,\Omega_j)\r]^2\,
T^2/2}}{\mathrm{e}^{\beta\, q}-1}\biggr\},
\end{eqnarray}
which is essentially the transition probability rate of a single detector that
we encountered earlier [cf. Eq~\eqref{eq:response-fn-finite-thermal}].
For instance,  the transition probability rate~$R_{11}^{T}(\cEb_1)$ is given by 
the expression~\eqref{eq:response-fn-finite-thermal} with $(v, \gamma,\cEb,\bar{\beta},
\bar{T})$ replaced by $(v_1,\gamma_{1},\cEb_{1},\bar{\beta}_1, \bar{T}_1)$.
In terms of the dimensionless variables $\bar{\gamma}$ and $\bar{\Omega}$ 
we had introduced, we find that the transition probability rate~$R_{22}^{T}(\cEb_1)$
can be written as 
\begin{eqnarray}\label{eq:R22-FT-TB}
R_{22}^{T}(\cEb_1) 
&=& \f{\sqrt{2\,\pi}\;\bar{T}_{1}}{4\,\pi\,\gamma_{1}}\,
\int_{0}^{\infty} \d x_{1}\,\sum_{m=-\infty}^{\infty}\,
J^2_{m}(x_{1}\,v_{2}/\bar{\Omega})\nn\\
& & \times\,\biggl\{\frac{\mathrm{e}^{-\l[\cEb_{1}+\bar{\gamma}\,
(x_1-m\,\bar{\Omega})\r]^2\,
\bar{T}_{1}^2/2}}{1-\mathrm{e}^{-\bar{\beta}_{1}\, x_{1}}}\nn\\
& & +\f{\mathrm{e}^{-\l[\cEb_{1}-\bar{\gamma}\,(x_1-m\,\bar{\Omega})\r]^2\,
\bar{T}_{1}^2/2}}{\mathrm{e}^{\bar{\beta}_{1}\, x_{1}}-1}\biggr\}.
\end{eqnarray}
Again, we can numerically compute the integral over the variable~$x_1$ and 
carry out the sum over~$m$.

\begin{figure}[!t]
\centering
\includegraphics[width=8.50cm]{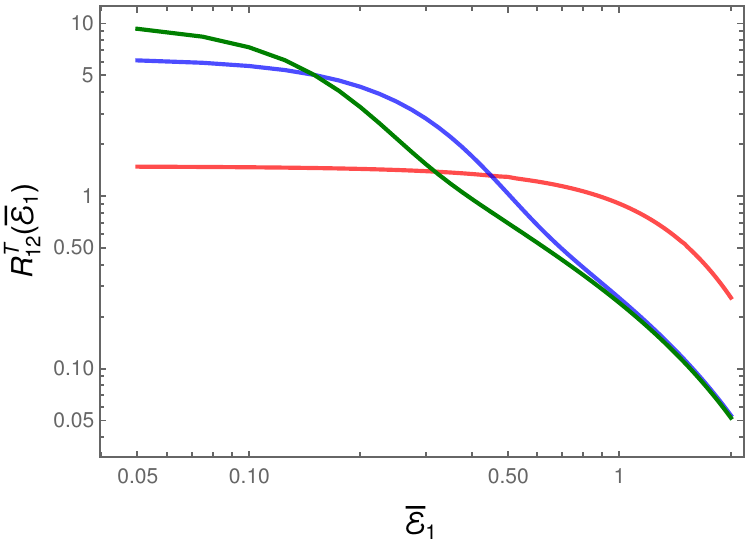}
\caption{The cross transition probability rate $R^{T}_{12}(\cEb_{1})$
of the two entangled detectors moving on circular trajectories and immersed 
in a thermal bath has been plotted as a function of the dimensionless energy 
gap~$\cEb_{1}$ for the case wherein the detectors remain switched on for a 
finite time interval.
As in the previous two figures, we have set $v_{1}=v_{2}=0.5$ (so that $\bar{\gamma}
=1$) and $\bar{\Omega}=5$.
We have plotted the results for $\bar{T}_{1}=(1,5,10)$ (in red, blue 
and green) and $\bar{\beta}_{1}=1$.
Note that the cross transition probability rate behaves in a manner similar to 
the auto transition probability rate we had plotted earlier (see 
Fig.~\ref{fig:RT11-fT-Mink-Vv}).}\label{fig:R12-FiniteT-Th-VEb1}
\end{figure}

When $j=1$ and $l=2$, the transition probability rate $R_{12}^{T}(\cEb_1)$ 
can be expressed as follows:
\begin{widetext}
\begin{eqnarray}\label{eq:R12-FT-TB}
R_{12}^{T}(\cEb_1) 
&=& \frac{\sqrt{2\,\pi}\,\bar{T}_{1}}{4\,\pi\,\gamma_{1}}\, 
\int_{0}^{\infty}\d x_{1} \sum_{m=-\infty}^{\infty}J_{m}(x_{1}\, v_{1})\,
J_{m}(x_{1}\, v_{2}/\bar{\Omega})\nn\\
& & \times\,\biggl(\f{1}{1-\mathrm{e}^{-\bar{\beta}_1\, x_1}}\, 
\mathrm{e}^{-\l\{\l[(1-\bar{\gamma})\,x_1-m\,(1-\bar{\gamma}\,\bar{\Omega})\r]^2 
+ \l[(1+\bar{\gamma})\,x_1-m\,(1+\bar{\gamma}\,\bar{\Omega})+2\,\cEb_{1}\r]^2\r\}
\bar{T}_1^2/8}\nn\\
& & +\f{1}{\mathrm{e}^{\bar{\beta}_1\, x_1}-1}\, 
\mathrm{e}^{-\l\{\l[(1-\bar{\gamma})\,x_1-m\,(1-\bar{\gamma}\,\bar{\Omega})\r]^2 
+ \l[(1+\bar{\gamma})\,x_1-m\,(1+\bar{\gamma}\,\bar{\Omega})-2\,\cEb_{1}\r]^2\r\}
\bar{T}_1^2/8}\biggr)=R_{21}^{T}(\cEb_1).
\end{eqnarray}
\end{widetext}
This expression also reduces to $R_{11}^{T}(\mathcal{E})$ when $\bar{\gamma}
= 1 = \bar{\Omega}$ and $v_{2} = v_{1}$.
As in the case of the response in the Minkowski vacuum, in App.~\ref{app:tpr-ft-fe},
we have provided the complete expressions for the auto and cross transition probability
rates of the rotating detectors in a thermal bath.
Moreover, we carry out the integral and the sum over the domain we had indicated 
earlier (in the caption of Fig.~\ref{fig:RT11-fT-Mink-Vv}).
In Fig.~\ref{fig:R12-FiniteT-Th-VEb1} we have plotted the cross transition 
probability rate $R_{12}(\cEb_{1})$ for different values of the dimensionless
time $\bar{T}_1$ and a fixed value of the dimensionless
inverse temperature $\bar{\beta}_1$.
Broadly, the cross transition probability rate exhibits the same behavior as 
the auto transition probability rate we had encountered earlier [cf. 
Fig.~\ref{fig:RT11-fT-Mink-Vv}].


\section{Summary and discussion}\label{sec:sd}

In this section, we shall summarize the results we have obtained and conclude 
with a broader discussion.


\subsection{Summary}

In the previous section, we had calculated the auto and the cross transition
probability rates of the two entangled detectors that are moving on circular
trajectories.
The auto transition probability rates of the detectors are evidently the same 
as the response of the single detectors we had discussed initially (in
Sec.~\ref{sec:rrd-mv-tb}).
Note that the complete transition probability of the entangled detectors is 
given by the expression~\eqref{eq:Transition-prob}, which involves contributions
from the auto and cross transition probabilities.
Let us now discuss the complete probability rates for 
transitions from the symmetric and anti-symmetric Bell states
to the excited state of the two entangled detectors.

Recall that the transition amplitude of the monopole operator is given by
$m_{j}^{\omega\bar{\omega}} = \langle \bar{\omega} \vert \hat{m}_{j}(0)\vert\omega\rangle$, 
with $\hat{m}_j(0)$ being defined in Eq.~\eqref{eq:mo}.
As we had mentioned, for a transition from the symmetric or the anti-symmetric 
Bell states (i.e. from $\vert s\rangle$ or $\vert a\rangle)$ to the collective 
excited state (i.e. $\vert e\rangle)$, the transition amplitudes of the 
monopole operator are found to be $m_{1}^{se} = m_{2}^{se} = 1/\sqrt{2}$ and 
$m_{1}^{ae} = -m_{2}^{ae}= -1/\sqrt{2}$. 
Due to this reason, the corresponding transition probability~\eqref{eq:Transition-prob} 
will contain an overall factor of~$1/2$, apart from the factor of~$\mu^2$ that arises
due to the strength of the coupling between the detectors and the scalar field.
Since the overall factor $\mu^2/2$ does not depend on either the trajectory of the 
detector or the state of the field, we shall drop the quantity or, equivalently, 
consider the total transition probability rate, say, 
$\mathcal{R}^{\tilde{T}}_{\omega\bar{\omega}}(\cE)$, to be given by 
\begin{eqnarray}\label{eq:transition-rate-ase}
\mathcal{R}^{\tilde{T}}_{\omega\bar{\omega}}(\cE) = \f{2}{\mu^2}\, 
\f{\Gamma_{|\omega\rangle\to|\bar{\omega}\rangle}(\cE)}{\tilde{T}},
\end{eqnarray}
where $\tilde{T}=\sqrt{(\pi/2)}\,T$ in the case of detectors that are switched
on for a finite duration through the Gaussian switching functions and $\tilde{T}
=\lim_{T\to \infty} T$ in the case of detectors that remain switched on forever.
Therefore, the total transition probability rates from the symmetric and 
anti-symmetric Bell states to the collective excited state, referred to by  
the subscripts `$se$' and `$ae$', respectively, can be expressed as
\begin{subequations}
\begin{eqnarray}\label{eq:transition-rate-ase-IfT}
\mathcal{R}_{se}^{\tilde{T}}(\cE) \!\!&=&\!\! R_{11}^{\tilde{T}}(\mathcal{E})
+ R_{22}^{\tilde{T}}(\mathcal{E})+\l[R_{12}^{\tilde{T}}(\mathcal{E})
+R_{21}^{\tilde{T}}(\mathcal{E})\r],\\
\mathcal{R}_{ae}^{\tilde{T}}(\cE) \!\!&=&\!\! R_{11}^{\tilde{T}}(\mathcal{E})
+ R_{22}^{\tilde{T}}(\mathcal{E})
-\l[R_{12}^{\tilde{T}}(\mathcal{E})+R_{21}^{\tilde{T}}(\mathcal{E})\r].\qquad\quad
\end{eqnarray}
\end{subequations}
Note that, in these expressions, for convenience, we have used the notation introduced
above, viz. that $R_{jl}^{\tilde{T}}(\mathcal{E})$ denotes the auto or cross transition 
probability rate of the detectors switched on for a finite or infinite time interval.

\begin{figure*}
\centering
\includegraphics[width=8.50cm]{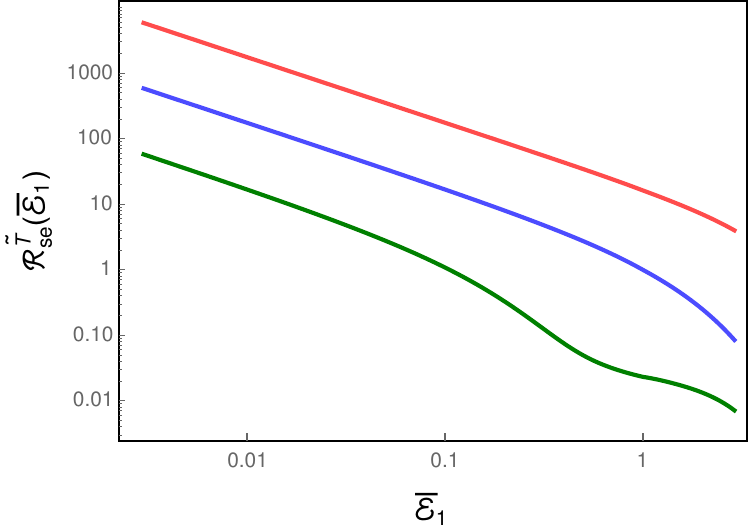}
\hskip 10pt
\includegraphics[width=8.50cm]{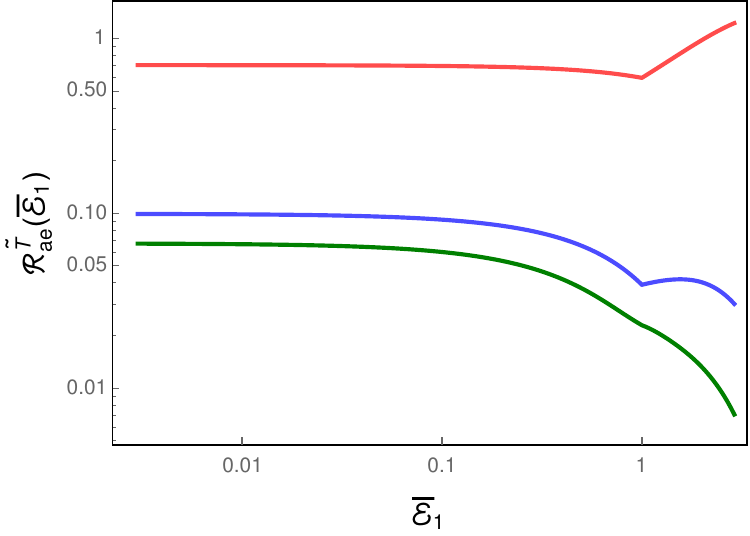}
\caption{The total transition probability rates $\mathcal{R}^{\tilde{T}}_{se}(\cEb_1)$ 
(on the left) and $\mathcal{R}^{\tilde{T}}_{ae}(\cEb_1)$ (on the right) of the two 
entangled, rotating detectors in a thermal bath have been plotted as 
functions of~$\cEb_{1}$ for the case wherein the detectors remain 
switched on forever.
We have set $v_{1}=v_{2}=0.5$, corresponding to $\bar{\gamma}=1$, and 
have chosen $\bar{\Omega}=5$, as we have done in the earlier figures.
We have plotted the results for three different values of the dimensionless 
inverse temperature, viz. $\bar{\beta}_{1}=(0.1,1,10)$ (in red, blue and green).
Note that, for the values of the parameters we have worked with, the total 
rate $\mathcal{R}^{\tilde{T}}_{se}(\cEb_1)$ is a factor of $10^3$ higher at
small energies than the rate $\mathcal{R}^{\tilde{T}}_{ae}(\cEb_1)$.}
\label{fig:Ysae-vs-DE-InFiniteT-thermal}
\end{figure*}

Let us first consider the case wherein the detectors are switched on for infinite 
duration.
When the scalar field is assumed to be in the Minkowski vacuum, in the situations
wherein $\gamma_1=\gamma_2$ and $\Omega_1\ne \Omega_2$ that we had focused on, the 
cross transition probability rates $R_{12}(\mathcal{E})$ and $R_{21}(\mathcal{E})$ 
vanish.
This implies that the total transition probability rates $\mathcal{R}_{se}^{\tilde{T}}(\cE)$ 
and $\mathcal{R}_{ae}^{\tilde{T}}(\cE)$ will be equal and both the rates can be entirely 
expressed in terms of the auto transition probability rates~$R_{11}(\mathcal{E})$ 
and~$R_{22}(\cE)$. 
As a result, the rates $\mathcal{R}_{se}^{\tilde{T}}(\cE)$ and
$\mathcal{R}_{ae}^{\tilde{T}}(\cEb)$ can be expected to be similar to that of, 
say, $R_{11}(\cE)$ (in this regard, see Fig.~\ref{fig:srd}).
When the detectors are assumed to be immersed in a thermal bath of quanta associated
with the scalar  field, we had found that, for $\gamma_1=\gamma_2$ and $\Omega_1\ne
\Omega_2$, the cross transition probability rates $R_{12}(\cE)$ and $R_{21}(\cE)$ 
prove to be non-zero.
Consequently, the total transition probability rates $\mathcal{R}_{se}^{\tilde{T}}(\cE)$ 
and $\mathcal{R}_{ae}^{\tilde{T}}(\cEb)$ can be expected to be different.
This is evident from Fig.~\ref{fig:Ysae-vs-DE-InFiniteT-thermal} where we have 
presented these total transition probability rates.
Note that the total rate $\mathcal{R}_{se}^{\tilde{T}}(\cEb_1)$ is about $10^{3}$ 
times larger in magnitude than the rate $\mathcal{R}_{se}^{\tilde{T}}(\cEb_1)$
for suitably small energies (in fact, for $\cEb_{1} \lesssim 0.01)$.
These findings can provide, in principle, observable distinction in the radiative 
processes of entangled detectors between the Minkowski vacuum and a thermal bath.

\begin{figure*}
\centering
\includegraphics[width=8.50cm]{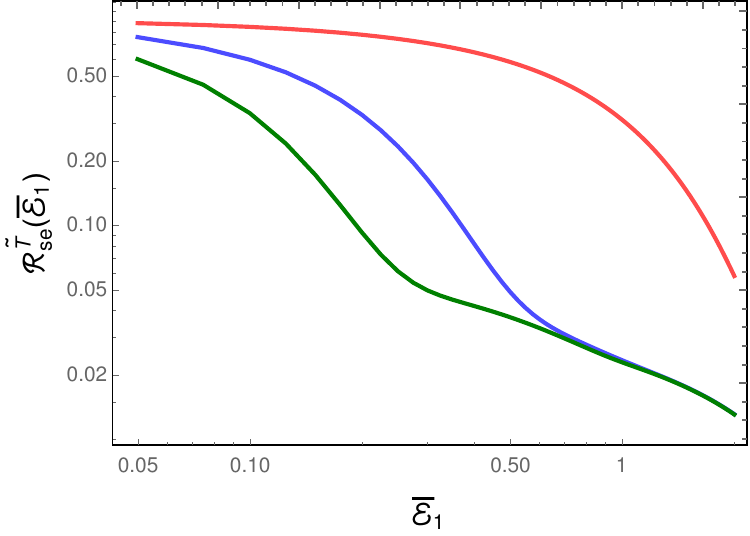}
\hskip 10pt
\includegraphics[width=8.50cm]{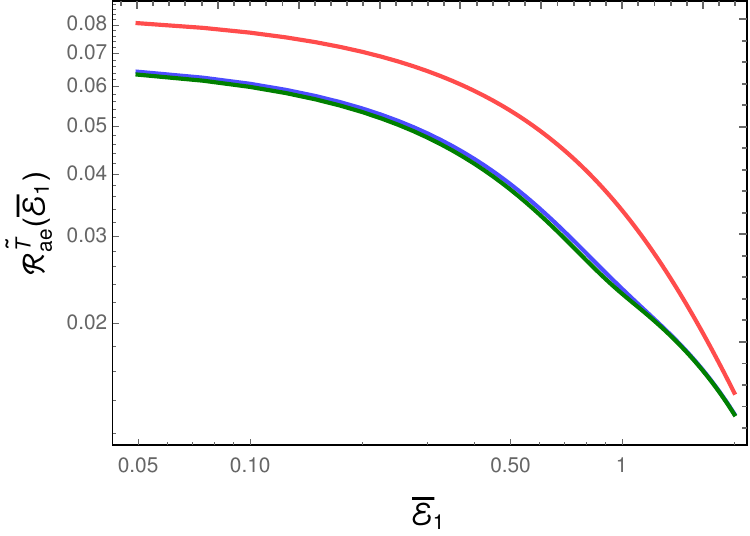}
\vskip 10pt
\includegraphics[width=8.50cm]{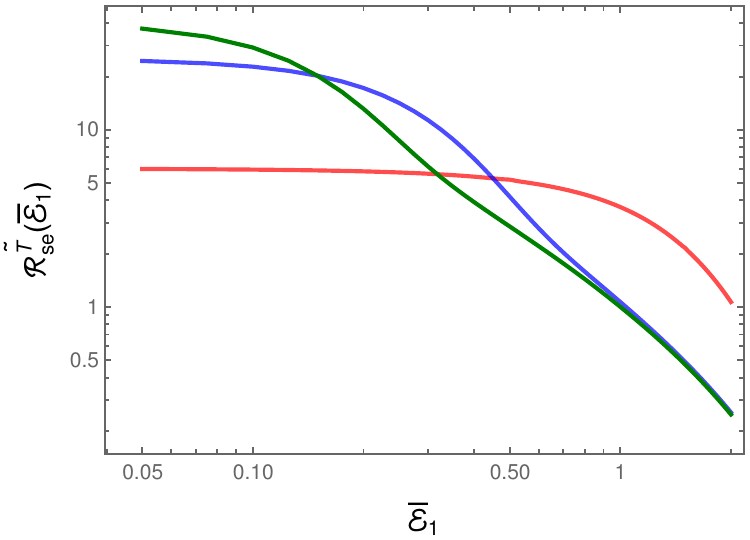}
\hskip 10pt
\includegraphics[width=8.50cm]{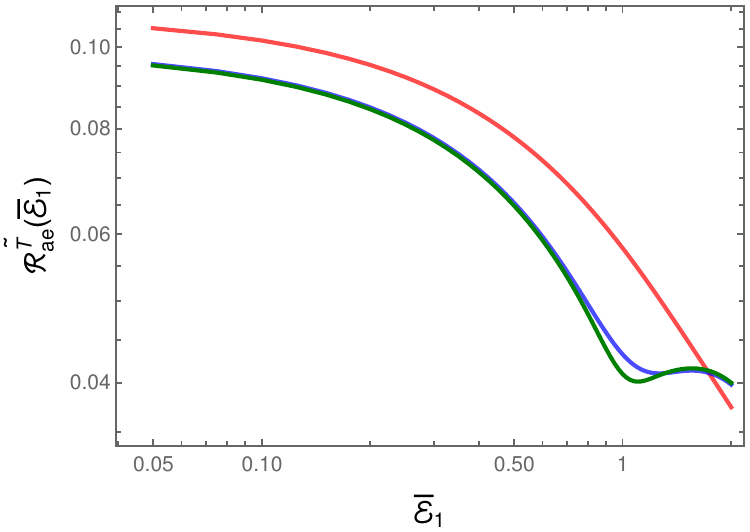}
\caption{The total transition probability rates $\mathcal{R}^{\tilde{T}}_{se}(\cEb_1)$ 
(on the left) and $\mathcal{R}^{\tilde{T}}_{ae}(\cEb_1)$ (on the right) of two entangled 
detectors moving on circular trajectories have been plotted as functions 
of~$\cEb_{1}$, when the detectors have been switched on for a finite time 
interval.
We have plotted the results in the Minkowski vacuum (on top) as well as in a 
thermal bath (at the bottom).
As before, we have set $v_{1}=v_{2}=0.5$ (corresponding to $\bar{\gamma}=1$) 
and have chosen $\bar{\Omega}=5$.
We have set the dimensionless inverse temperature of the thermal bath to be 
$\bar{\beta}_{1}=1$.
Moreover, as earlier, we have plotted the total rates for three different 
values of the dimensionless time interval, viz. $\bar{T}_{1}=(1,5,10)$ (in 
red, blue and green). 
Clearly, the total rate $\mathcal{R}^{\tilde{T}}_{se}(\cEb_1)$ 
is significantly higher than the rate $\mathcal{R}^{\tilde{T}}_{ae}(\cEb_1)$ due
to the interference effects.}\label{fig:Ysae-vs-DE-FiniteT-Minkowski}
\end{figure*}

Let us now discuss the cases wherein the detectors are switched on for a finite time
interval using the Gaussian switching functions.
In Fig.~\ref{fig:Ysae-vs-DE-FiniteT-Minkowski}, we have plotted the total transition
probability rates $\mathcal{R}_{se}^{\tilde{T}}(\cE)$ 
and $\mathcal{R}_{ae}^{\tilde{T}}(\cEb)$ of the entangled detectors in the Minkowski 
vacuum as well as the thermal bath.
We find that the total rate $\mathcal{R}_{se}^{\tilde{T}}(\cE)$ for a transition 
from the symmetric Bell state to the excited state is nearly $10$ times than the 
total rate $\mathcal{R}_{ae}^{\tilde{T}}(\cEb)$ from the asymmetric Bell state, 
due to the constructive and destructive interference we mentioned above.
Apart from this aspect, the total rates broadly exhibit characteristics that 
are similar to what we had encountered in  the auto and cross transition 
probability rates.


\subsection{Discussion}

In this work, we have examined the response of detectors that are moving
on circular trajectories in $(2+1)$~dimensional flat spacetime. 
As has been pointed out before (in this context, see, for example, 
Ref.~\cite{Bell:1982qr}), it seems more realistic and practical to 
consider detectors that are in motion on circular trajectories than 
detectors that are moving on uniformly accelerated trajectories.
We should mention that certain aspects of the response of entangled 
detectors that are in motion on circular trajectories have been studied 
earlier in the literature (see, for instance, Ref.~\cite{Costa:2020aqa}).
We believe that there are many interesting aspects of the rotating detectors 
that we have uncovered.
To begin with, we find that, in the case of two entangled, rotating detectors, 
the cross transition probability rates can be comparable to the auto transition 
probability rates of the individual detectors (cf. Figs.~\ref{fig:R12-InfT-Th-VEb1},
\ref{fig:R12-FiniteT-Mink-VEb1}, \ref{fig:R12-FiniteT-Th-VEb1}).
Second, when the detectors are switched on for infinite duration (in both 
single and entangled cases), the transition probability rate of the rotating
detectors in the Minkowski vacuum and a thermal bath are higher at smaller 
values of the energy gap of the detectors, higher values of their velocity 
and higher values of the temperature (cf. Fig.~\ref{fig:srd}).
Third, in the Minkowski vacuum, interestingly, we find that the transition
probability rates of the detectors are higher when they are switched on for 
a shorter duration (cf. Fig.~\ref{fig:RT11-fT-Mink-Vv}).
Though, at first, this result may seem counter-intuitive, it can be
interpreted as a manifestation of the energy-time uncertainty principle.
The shorter the interval of time that the detector remains switched on, the 
larger can be the energy of the virtual quanta that are available to excite 
the detector. 
Fourth, in a thermal bath, when the detectors are switched on for a finite 
time interval, we  observe that the transition probability rate is higher 
for smaller intervals of time only when the temperature of the thermal bath
is lower or the energy gap of the detectors is higher.
In fact, we observe that the behavior can be reversed at higher temperatures
and smaller energy gaps (cf. Figs.~\ref{fig:RT11-fT-Mink-Vv} 
and~\ref{fig:R11-fT-Th-Vbeta}).
Fifth, from Eq.~\eqref{eq:response-fn-infinite-thermal2} and the related 
discussions in Sec.~\ref{sec:rrd-mv-tb}, we identified a specific difference 
in the nature of the response of single Unruh-DeWitt detectors in a thermal bath,
while they are on circular trajectories, when compared to the accelerated case.
There is a single spontaneous excitation in the case of circular trajectories due 
to the motion, while the thermal bath contributes to the stimulated excitations. 
On the other hand, there are two independent, spontaneous excitations due to 
the motion and the thermal bath in the accelerated case, in addition to the 
stimulated excitation due to the thermal bath.
Finally, as we had discussed in the previous subsection, due to constructive or 
destructive interference, the total transition probability rates from the symmetric 
and anti-symmetric Bell states to the collective excited state can be substantially 
different in a thermal bath or when they are switched on for a finite time interval 
in the Minkowski vacuum. 
We should mention that a similar behavior is also observed when one considers 
the de-excitation of the detector from the symmetric or the anti-symmetric Bell 
states to the ground state (in this regard, see the discussion in 
Refs.~\cite{FICEK2002369,Menezes:2015uaa}).

There are many further aspects of the rotating and entangled detectors that 
remain to be explored.
We need to urgently extend all our analysis to $(3+1)$-spacetime dimensions.
In $(3+1)$-spacetime dimensions, while we expect the results to be qualitatively 
similar to the $(2+1)$-dimensional case we have considered here, we can expect 
some quantitative differences.
Also, we have to examine whether two initially uncorrelated 
atomic detectors moving on circular trajectories can get entangled over time, a 
phenomenon that has been referred to as entanglement harvesting (in this regard, 
see Refs.~\cite{Reznik:2002fz, Martin-Martinez:2015qwa,
Koga:2018the, Koga:2019fqh, Barman:2021bbw, Barman:2021kwg}). 
In particular, we need to investigate entanglement harvesting in the presence 
of a thermal bath, with detectors switched on for infinite as well as finite
intervals of time (for previous studies in this context involving static and 
non-inertial detectors in a thermal bath, see 
Refs.~\cite{Brown:2013kia,Simidzija:2018ddw} 
and  Ref.~\cite{Barman:2021bbw}, respectively).
Moreover, it will be interesting to study the effects due to the presence
of boundaries~\cite{Davies:1996ks,Costa:2020aqa}.
Further, on the practical front, it is easier to set charged particles in
motion on circular trajectories, using, say, with the help of an external 
magnetic field.
If a charged particle (say, an ion) is to be used as a detector, then it
may emit classical synchrotron radiation as it moves along the circular
trajectories.
We need to understand the implications or effects of the synchrotron
radiation for the detection of quanta emitted or absorbed by the detector
due to the quantum phenomena we are investigating.
We are presently working on these issues.


\begin{acknowledgments}
SB would like to thank the Science and Engineering Research Board, 
Government of India, for supporting this work through the National 
Postdoctoral Fellowship~PDF/2022/000428.
\end{acknowledgments}


\appendix

\section{Wightman function in polar coordinates}\label{app:Evln-GreensFn}

In $(2+1)$-dimensional flat spacetime, when working in the Cartesian 
coordinates ${\bm x}=(x,y)$, the Wightman function associated with a 
massive, minimally coupled, scalar field can be written as (in this 
regard, see, for instance, Ref.~\cite{book:Birrell}) 
\begin{eqnarray}\label{eq:AppA-GnFn-1}
G^+(\tx,\tx') = \int\frac{\d^2{\bm k}}{(2\,\pi)^2\, (2\,\omega)}\,
\mathrm{e}^{-i\,\omega\,(t-t')+i\,{\bm k}\cdot({\bm x}-{\bm x}')},
\end{eqnarray}
where $\omega = (\vert {\bm k}\vert^2+\mu^2)^{1/2}$, and $\mu$ denotes 
the mass of the field.
Let us write both the wave vector~${\bm k}$ and the position vector~${\bm x}$
in terms of the corresponding polar coordinates, say, $(q,\alpha)$ and $(\rho,
\phi)$, as follows:
\begin{subequations}\label{eq:AppA-WvVec-transformation}
\begin{eqnarray}
k_x &=& q\,\cos{\alpha},\quad\; k_y = q\,\sin{\alpha},\\
x &=& \rho\,\cos{\phi},\quad\;\;\, y = \rho\,\sin{\phi},\\
x' &=& \rho'\,\cos{\phi'},\quad y' = \rho'\,\sin{\phi'}.
\end{eqnarray}
\end{subequations}
In such a case, the Wightman function~\eqref{eq:AppA-GnFn-1} can be 
expressed as
\begin{eqnarray}\label{eq:AppA-GnFn-2}
G^+(\tx,\tx') 
&=& \int_{0}^{\infty}\f{\d q\, q}{(2\,\pi)\,(2\,\omega)}
\int_{0}^{2\,\pi} \f{\d\alpha}{2\,\pi}\,\mathrm{e}^{-i\,\omega\,(t-t')}\nn\\
& &\times\,\mathrm{e}^{i\,q\,\l[\rho\,\cos(\phi-\alpha)
-\rho'\,\cos (\phi'-\alpha)\r]}.
\end{eqnarray}
If we now use the following identity (known as the \emph{Jacobi–Anger identity};\/
in this context, see, for instance, Ref.~\cite{garfken67:math})
\begin{eqnarray}\label{eq:AppA-rel-Arfken}
\mathrm{e}^{i\,z\,\cos{\phi}} 
= \sum_{m=-\infty}^{\infty}i^m\,J_{m}(z)\,\mathrm{e}^{i\,m\,\phi},
\end{eqnarray}
where $J_m(z)$ are the Bessel functions, then the Wightman function can be
written as
\begin{eqnarray}\label{eq:AppA-GnFn-3}
G^+(\tx,\tx') 
&=& \int_{0}^{\infty}\f{\d q\,q}{(2\,\pi)\,(2\,\omega)}\,\int_{0}^{2\pi}\,
\f{\d \alpha}{2\,\pi}\,\mathrm{e}^{-i\,\omega\,(t-t')}\nn\\
& &\times\,\sum_{m=-\infty}^{\infty}i^m\,J_{m}(q\,\rho)\,
\mathrm{e}^{i\,m\,(\phi-\alpha)}\,\nn\\
& &\times\,\sum_{m'=-\infty}^{\infty}i^{m'}\,J_{m'}(-q\,\rho')\,
\mathrm{e}^{i\,m'\,(\phi'-\alpha)}\nn\\
&=& \int_{0}^{\infty}\f{\d q\, q}{4\,\pi\,\omega}\,
\mathrm{e}^{-i\,\omega\,(t-t')}\nn\\
& &\times\,\sum_{m=-\infty}^{\infty}J_{m}(q\,\rho)\,
J_{-m}(-q\,\rho')\,\mathrm{e}^{i\,m\,(\phi-\phi')}.\nn\\
\end{eqnarray}
We should mention that, to arrive at the final equality, we have used the 
relation
\begin{equation}\label{eq:AppA-rel-2}
\int_{0}^{2\,\pi} \d \alpha\,
\mathrm{e}^{-i\,(m+m')\,\alpha} = (2\,\pi)\,\delta_{m,-m'},
\end{equation}
where $\delta_{n,n'}$ denotes the Kronecker delta. 
On using the identity $J_{-m}(-q\,\rho') = J_{m}(q\,\rho')$, in the case
of a massless field (i.e. when $\mu=0$ so that $\omega=q$), we can arrive
at the expression~\eqref{eq:Greenfn-1detector} for the Wightman function 
we have mentioned earlier.

Similarly, when working in the Cartesian coordinates, the Wightman function for 
the scalar field at a finite temperature in $(2+1)$ spacetime dimensions can be 
easily obtained to be (see, for instance, Ref.~\cite{book:Birrell,Kolekar:2013hra, 
Chowdhury:2019set})
\begin{eqnarray}\label{eq:AppA-GnFn-2}
G_\beta^+(\tx,\tx') 
&=& \int\frac{\d^2{\bm k}}{(2\,\pi)^2\, (2\,\omega)}\,
\biggl\{\f{\mathrm{e}^{-i\,\l[\omega\,(t-t')-{\bm k}\cdot({\bm x}-{\bm x}')\r]}}
{1-\mathrm{e}^{-\beta\,\omega}}\nn\\
& &+\f{\mathrm{e}^{i\,\l[\omega\,(t-t')-{\bm k}\cdot({\bm x}-{\bm x}')\r]}}
{\mathrm{e}^{\beta\,\omega-1}}\biggr\}.
\end{eqnarray}
In the case of a massless field, upon carrying out the 
transformations~\eqref{eq:AppA-WvVec-transformation}, we can arrive at the
expression for the above Wightman function in terms of the polar coordinates 
(in both real and momentum space), which is the 
result~\eqref{eq:Greenfn-2p1-thermal} we have quoted earlier.


\section{Behavior of the transition probability rate as a function of velocity 
of the detector}\label{app:m}

\begin{figure}[!t]
\centering
\includegraphics[width=8.50cm]{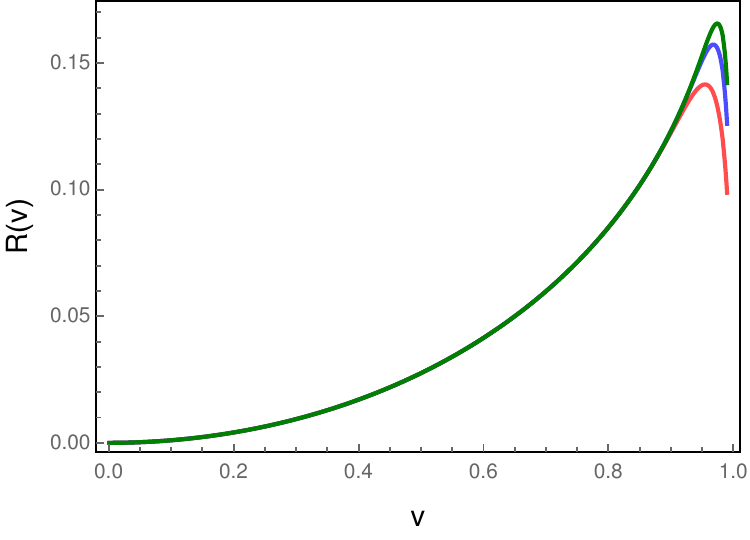}
\caption{The transition probability rate of the detector $R(\cEb,v)$ in 
the Minkowski vacuum, when it is moving on a circular trajectory and has been
switched on for infinite duration, has been plotted as a function of the 
velocity~$v$, when the sum over $m$ is carried out to larger and larger values,
viz. until $(m-\cEb)=(50, 100,150)$ (in red, blue and green, respectively).
We have set $\cEb=0.1$ in arriving at these plots.
Note that, for velocities close to unity, the peak in the transition probability
rates shifts towards higher values of the velocity as we sum until larger values
of~$m$.
This illustrates that care needs to be exercised when the sum over~$m$ is 
carried out.
In the plots we have presented earlier, at every stage, we have checked and 
confirmed that the terms we have ignored do not contribute significantly to
the transition probability rate of the detector.}\label{fig:RT11-fT-Mink-tb-Vv}
\end{figure}

In our discussion, barring in Fig.~\ref{fig:R11-fT-Th-Vbeta}, we have been 
primarily interested in computing the auto and cross transition probability 
rates of the detectors as a function of the energy gap~$\cE$, for given 
angular and linear velocities~$\Omega$ and~$v$ of the detector, inverse 
temperature~$\beta$ of the thermal bath and the time interval~$T$ for which 
the detector is switched on.
We had pointed out that the sums over $m$ which appear in the transition 
probability rates converge fairly quickly.
Specifically, we had mentioned that, for the parameters we have considered, it 
is adequate to evaluate the sum until $(m-\cEb) = 50$ and $(m-\cEb) = 10$ when
the detectors are switched on for infinite or a finite duration (in this regard, 
see the captions of Figs.~\ref{fig:srd} and~\ref{fig:RT11-fT-Mink-Vv}).
However, when the detector is switched on for infinite duration, as we had 
discussed in Sec.~\ref{sec:rrd-mv-tb} [see our discussion following 
Eq.~\eqref{eq:response-fn}], the convergence of the sum depends on the velocity 
of the detector.
We find that, for detector velocities very close to the velocity of light (say,
for $v\gtrsim 0.9$), it becomes necessary to evaluate the sum to larger values 
of~$m$.
To illustrate this point, in Fig.~\ref{fig:RT11-fT-Mink-tb-Vv}, we have plotted 
the transition probability rate of the rotating detector in the Minkowski vacuum
as a function of the linear velocity~$v$ of the detector.
We have fixed the value of the energy gap~$\cE$ in plotting the results and have 
assumed that the detector remains switched on forever.
Note that, for~$v\gtrsim 0.9$, the peak in the transition probability rate of
the detector shifts towards higher velocities as we sum to larger and larger 
values of~$m$.
In the results we have presented in all the earlier figures, we have ensured that, 
for the parameters we have worked with, summing to larger values of~$m$ does not 
significantly change the results we obtain.


\section{Behavior of the integrand in the finite time transition probability
rate}\label{app:convergence-int-x}

\begin{figure*}
\centering
\includegraphics[width=8.50cm]{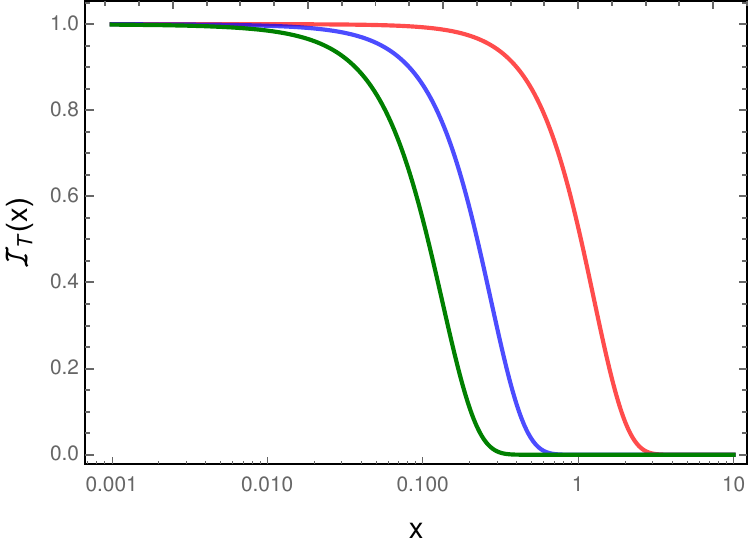}
\hskip 10pt
\includegraphics[width=8.50cm]{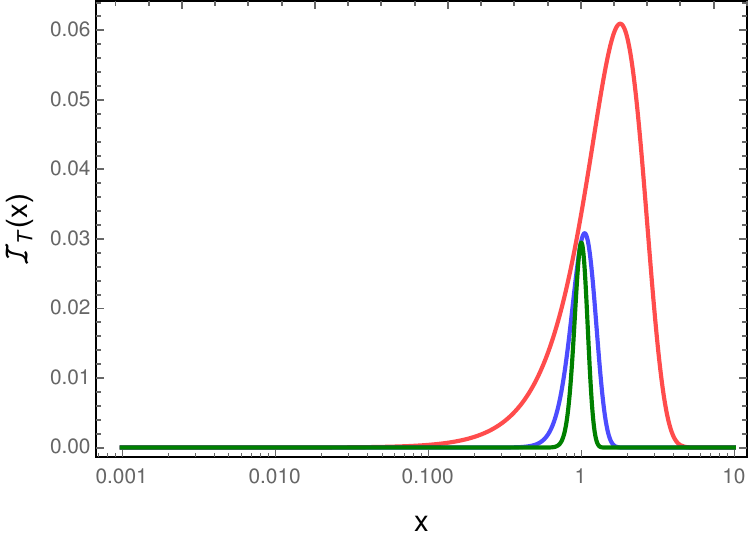}
\caption{The integrand $\mathcal{I}_T(x)$ 
[cf. Eq.~\eqref{eq:Integral-x-app} has been plotted as a function of the 
dimensionless variable~$x$ for $m=0$ (on the left) and $m=1$ (on the right).
We have chosen $v=0.5$ and $\cEb=0.01$, and have plotted the integrand for  
$\bar{T}=(1,5,10)$ (in red, blue, and green).
Clearly, the integrand is well behaved at small~$x$ and quickly dies down at 
large $x$, allowing us to efficiently compute the 
integrals.}\label{fig:convergence-int-x}
\end{figure*}

Recall that, in the case of detectors that are switched on for a finite time
interval, we have to carry out an integral over $x$, apart from summing 
over~$m$.
Immediately after Eq.~\eqref{eq:response-fn-finite-Minkowski}, we had discussed 
the behavior of the integrals at large and small values of~$x$.  
In this appendix, we shall briefly illustrate the behavior of the integrands
that are encountered when evaluating the response of the detector in the 
Minkowski vacuum.
Note that the integrands in this case are of the following form:
\begin{eqnarray}\label{eq:Integral-x-app}
\mathcal{I}_T(x) 
&=& J^2_{m}(x\,v)\, \mathrm{e}^{-\l[(x^2/2)+x\,\cEb\r]\,\bar{T}^2}\,
\mathrm{e}^{-m^2\,\bar{T}^2/2}\nn\\
& &\times\,\mathrm{cosh}\l[m\,(x+\cEb)\,\bar{T}^2\r].
\end{eqnarray}
In Fig.~\ref{fig:convergence-int-x}, we have plotted this integrand for $m=0$
and $m=1$ and a few different values of the dimensionless time interval~$\bar{T}$,
assuming fixed values for the velocity~$v$ and the dimensionless energy gap~$\cEb$.
It is evident that the integrands are well behaved and, in particular, they
decrease rapidly at large~$x$ due to the $\mathrm{e}^{-x^2\,\bar{T}^2/2}$
factor.
Moreover, for $m>0$, the factor $\mathrm{e}^{-m^2\,\bar{T}^2/2}$
also suppresses the overall amplitude of the integrand.
Such a rapid decrease allows us to quickly compute the integrals involved.
As we had mentioned, for the values of the parameters we work with, we find
that it is adequate to integrate up to $x=10^2$ (see the caption of
Fig.~\ref{fig:RT11-fT-Mink-Vv}) and we have also checked that the results we 
obtain are not altered if we increase the upper limit.


\begin{widetext}
\section{Transition probability rates of entangled detectors for finite 
duration}\label{app:tpr-ft-fe}

In this appendix, we shall provide explicit expressions for the transition 
probability rates of the entangled detectors that are moving on circular
trajectories and are interacting with the scalar field for a finite time 
interval.
It is these expressions that we eventually use to numerically compute the 
transition probability rates.

Let us first consider the case of the Minkowski vacuum. 
We find that the expressions~\eqref{eq:Rjj-FTM-S1} and~\eqref{eq:R22-FTM-S1}
can be utilized to write $R_{11}^{T}(\cEb_1)$ and $R_{22}^{T}(\cEb_1)$ as
\begin{subequations}
\begin{eqnarray}
R_{11}^{T}(\cEb_1) 
&=& \f{\sqrt{2\,\pi}\;\bar{T}_{1}}{4\,\pi\,\gamma_{1}}\,
\mathrm{e}^{-\cEb_{1}^2\,\bar{T}_{1}^2/2}\,
\Biggl\{\int_{0}^{\infty} \d x_{1}\,J^2_{0}(x_{1}\,v_{1})\,
\mathrm{e}^{-\l[(x_{1}^2/2)+x_{1}\,\cEb_{1}\r]\,\bar{T}_{1}^2}\nn\\
& &+\,2\,\sum_{m=1}^{\infty}\,  \mathrm{e}^{-m^2\,\bar{T}_{1}^2/2}
\int_{0}^{\infty} \d x_{1}\, J^2_{m}(x_{1}\,v_{1})\,
\mathrm{e}^{-\l[(x_{1}^2/2)+x_{1}\,\cEb_{1}\r]\,\bar{T}_{1}^2}\,
\mathrm{cosh}\l[m\,(x_{1}+\cEb_{1})\,\bar{T}_{1}^2\r]\Biggr\}\,,\\
R_{22}^{T}(\mathcal{E}) 
&=& \f{\sqrt{2\,\pi}\;\bar{T}_{1}}{4\,\pi\,\gamma_{1}}\,
\mathrm{e}^{-\cEb_{1}^2\,\bar{T}_{1}^2/2}\,
\Biggl\{\int_{0}^{\infty} \d x_{1}\,J^2_{0}\l(x_{1}\,v_{2}/\bar{\Omega}\r)\,
\mathrm{e}^{-\l[(x_1^2/2)+x_{1}\,\cEb_{1}/\bar{\gamma}\r]\,
\bar{\gamma}^2\,\bar{T}_{1}^2}\nn\\
& &+\,2\,\sum_{m=1}^{\infty}\,
\mathrm{e}^{-m^2\,\bar{\gamma}^2\,\bar{\Omega}^2\,\bar{T}_{1}^2/2}\,
\int_{0}^{\infty} \d x_{1}\, J^2_{m}\l(x_{1} v_{2}/\bar{\Omega}\r)\,
\mathrm{e}^{-\l[(x_{1}^2/2)
+ x_{1}\, \cEb_{1}/\bar{\gamma}\r]\,\bar{\gamma}^2\,\bar{T}_{1}^2}\nn\\
& &\times\,\mathrm{cosh}\l[m\,\l(x_{1}+\cEb_{1}/\bar{\gamma}\r)\,\bar{\gamma}^2\,
\bar{\Omega}\,\bar{T}_{1}^2\r]\Biggr\}.\qquad\quad
\end{eqnarray}
\end{subequations}
Similarly, the expression~\eqref{eq:R12-FTM-S1} for the cross transition probability 
rate $R_{12}^{T}(\cEb_1)$ in the Minkowski vacuum can be written as
\begin{eqnarray}
R_{12}^{T}(\cEb_1) 
&=& \frac{\sqrt{2\,\pi}\,\bar{T}_{1}}{4\,\pi\,\gamma_{1}}\,
\mathrm{e}^{-\cEb_{1}^2\,\bar{T}_{1}^2/2}\,
\Biggl(\int_{0}^{\infty} \d x_{1}\,J_{0}(x_{1}\,v_{1})\,
J_{0}(x_{1}\,v_{2}/\bar{\Omega})\,
\mathrm{e}^{-\l[(1+\bar{\gamma}^2)\,x_{1}^2
+2\,(1+\bar{\gamma})\,x_{1}\,\cEb_{1}\r]\,\bar{T}_{1}^2/4}\nn\\
& &+\,2\,\sum_{m=1}^{\infty}\,  
\mathrm{e}^{-m^2\,(1+\bar{\gamma}^2\,\bar{\Omega}^2)\,\bar{T}_{1}^2/4}
\int_{0}^{\infty} \d x_{1}\,  J_{m}(x_{1}\,v_{1})\,
J_{m}(x_{1}\,v_{2}/\bar{\Omega})\, 
\mathrm{e}^{-\l[(1+\bar{\gamma}^2)\,x_1^2
+2\,(1+\bar{\gamma})\,x_{1}\,\cEb_{1}\r]\,\bar{T}_{1}^2/4}\nn\\
& &\times\, \mathrm{cosh}\l\{m\,\l[(1+\bar{\gamma}^2\,\bar{\Omega})\,x_1
+(1+\bar{\gamma}\,\bar{\Omega})\,\cEb_{1}\r]\,\bar{T}_{1}^2/2\r\}\Biggr)
= R_{21}^{T}(\cEb_1).
\end{eqnarray}

The expressions~\eqref{eq:Rjj-FT-TB} and~\eqref{eq:R22-FT-TB} can be used
to write the auto transition probability rates of the rotating detectors 
in a thermal bath as
\begin{subequations}
\begin{eqnarray}\label{eq:appn-Auto-R11-R22-FiniteT-thermal}
R_{11}^{T}(\cEb_1) 
&=& \f{\sqrt{2\,\pi}\;\bar{T}_{1}}{4\,\pi\,\gamma_{1}}\,
\mathrm{e}^{-\cEb_{1}^2\,\bar{T}_{1}^2/2}\,
\Biggl\{\int_{0}^{\infty} \d x_{1}\,J^2_{0}(x_{1}\,v_{1})\,
\l[\frac{\mathrm{e}^{-\l[(x_{1}^2/2)+x_{1}\,\cEb_{1}\r]\,\bar{T}_{1}^2}}{1
-\mathrm{e}^{-\bar{\beta}_{1}\,x_{1}}}
+\f{\mathrm{e}^{-\l[(x_{1}^2/2)-x_{1}\,\cEb_{1}\r]\,
\bar{T}_{1}^2}}{\mathrm{e}^{\bar{\beta}_{1}\,x_{1}}-1}\,\r]\nn\\
& & +\,2\,\sum_{m=1}^{\infty}\,  \mathrm{e}^{-m^2\,\bar{T}_{1}^2/2}
\int_{0}^{\infty} \d x_{1}\, J^2_{m}(x_{1}\,v_{1})\,
\biggl[\f{\mathrm{e}^{-\l[(x_{1}^2/2)+x_{1}\,\cEb_{1}\r]\,\bar{T}_{1}^2}}{1
-\mathrm{e}^{-\bar{\beta}_{1}\,x_{1}}}\,\mathrm{cosh}\l[m\,(x_{1}+\cEb_{1})\,
\bar{T}_{1}^2\r]\nn\\
& & +\,\frac{\mathrm{e}^{-\l[(x_{1}^2/2)-x_{1}\,\cEb_{1}\r]\,
\bar{T}_{1}^2}}{\mathrm{e}^{\bar{\beta}_{1}\,x_{1}}-1}\,
\mathrm{cosh}\l[m\,(x_{1}-\cEb_{1})\,\bar{T}_{1}^2\r]\Biggr\},\\
R_{22}^{T}(\cEb_1) 
&=& \f{\sqrt{2\,\pi}\;\bar{T}_{1}}{4\,\pi\gamma_{1}}\,
\mathrm{e}^{-\cEb_{1}^2\,\bar{T}_{1}^2/2}\,\Biggl(\int_{0}^{\infty} \d x_{1}\,
J^2_{0}\l(\frac{x_{1}\,v_{2}}{\bar{\Omega}}\r)\,
\l[\f{\mathrm{e}^{-\l[(x_{1}^2/2)+x_{1}\,\cEb_{1}/\bar{\gamma}\r]\,
\bar{\gamma}^2\,\bar{T}_{1}^2}}{1-\mathrm{e}^{-\bar{\beta}_{1}\,x_{1}}}
+\frac{\mathrm{e}^{-\l[(x_{1}^2/2)-x_{1}\,\cEb_{1}/\bar{\gamma}\r]\,
\bar{\gamma}^2\,\bar{T}_{1}^2}}{\mathrm{e}^{\bar{\beta}_{1}\,x_{1}}-1}\r]\nn\\
& & +\,2\,\sum_{m=1}^{\infty}\,  
\mathrm{e}^{-m^2\,\bar{\gamma}^2\,\bar{\Omega}^2\,\bar{T}_{1}^2/2}\,
\int_{0}^{\infty} \d x_{1}\, J^2_{m}\l(x_{1} v_{2}/\bar{\Omega}\r)\,
\biggl\{\f{\mathrm{e}^{-\l[(x_{1}^2/2)+x_{1}\,\cEb_{1}/\bar{\gamma}\r]\,
\bar{\gamma}^2\,\bar{T}_{1}^2}}{1-\mathrm{e}^{-\bar{\beta}_{1}\,x_{1}}}\, 
\mathrm{cosh}\l[m\, \l(x_{1}+\f{\cEb_{1}}{\bar{\gamma}}\r)\,
\bar{\gamma}^2\,\bar{\Omega}\,\bar{T}_{1}^2\r]\nn\\
& & +\,\f{\mathrm{e}^{-\l[(x_1^2/2)-x_1\,\cEb_{1}/\bar{\gamma}\r]\,
\bar{\gamma}^2\,\bar{T}_{1}^2}}{\mathrm{e}^{\bar{\beta}_{1}\,x_{1}}-1} 
\mathrm{cosh}\l[m \l(x_{1}-\frac{\cEb_{1}}{\bar{\gamma}}\r)\, 
\bar{\gamma}^2\,\bar{\Omega}\,\bar{T}_{1}^2\r]\biggr\}\Biggr).
\end{eqnarray}
\end{subequations}
Similarly,  upon using the expression~\eqref{eq:R12-FT-TB},  the corresponding 
cross transition transition probability rate $R_{12}^{T}(\cEb_1)$ can be 
written as
\begin{eqnarray}\label{eq:appn-R12-FiniteT-thermal}
R_{12}^{T}(\cEb_1) 
&=& \frac{\sqrt{2\,\pi}\;\bar{T}_{1}}{4\,\pi\,\gamma_1} \,
\mathrm{e}^{-\cEb_{1}^2\,\bar{T}_{1}^2/2}\,
\Biggl[\int_{0}^{\infty} \d x_{1}\,J_{0}(x_{1}\,v_{1})\,J_{0}(x_{1}\,v_{2}/\bar{\Omega})\,
\biggl[\f{\mathrm{e}^{-\l[(1+\bar{\gamma}^2)\,x_1^2
+2\,(1+\bar{\gamma})\,x_{1}\,\cEb_{1}\r]\,\bar{T}_{1}^2/4}}{1
-\mathrm{e}^{-\bar{\beta}_{1}\,x_1}}\,\nn\\
& & +\,\f{\mathrm{e}^{-\l[(1+\bar{\gamma}^2)\,x_1^2
-2\,(1+\bar{\gamma})\,x_{1}\,\cEb_{1}\r]\,
\bar{T}_{1}^2/4}}{\mathrm{e}^{\bar{\beta}_{1}\,x_{1}}-1}\biggr]\nn\\
& &+\,2\,\sum_{m=1}^{\infty}\,
\mathrm{e}^{-m^2\,(1+\bar{\gamma}^2\,\bar{\Omega}^2)\,\bar{T}_{1}^2/4}\,
\int_{0}^{\infty} \d x_1\,  J_{m}(x_{1}\,v_{1})\,J_{m}(x_{1}\,v_{2}/\bar{\Omega})\nn\\
& &\times\,\bigg(\f{\mathrm{e}^{-\l[(1+\bar{\gamma}^2)\,x_1^2
+2\,(1+\bar{\gamma})\,x_{1}\,\cEb_{1}\r]\,\bar{T}_{1}^2/4}}{1
-\mathrm{e}^{-\bar{\beta}_{1}\,x_{1}}}\, 
\mathrm{cosh}\l\{m\,\l[ (1+\bar{\gamma}^2\,\bar{\Omega})\,x_1
+(1+\bar{\gamma}\,\bar{\Omega})\,\cEb_{1}\r]\,\bar{T}_{1}^2/2\r\}\nn\\
& & +\, \f{\mathrm{e}^{-\l[(1+\bar{\gamma}^2)\,x_1^2
-2\,(1+\bar{\gamma})\,x_{1}\,\cEb_{1}\r]\,
\bar{T}_{1}^2/4}}{\mathrm{e}^{\bar{\beta}_{1}\,x_{1}}-1}\, 
\mathrm{cosh}\l\{m\,\l[(1+\bar{\gamma}^2\,\bar{\Omega})\,x_1
-(1+\bar{\gamma}\,\bar{\Omega})\,\cEb_{1}\r]\,\bar{T}_{1}^2/2\r\}\biggr)\Biggr]
= R_{21}^{T}(\cEb_1).
\end{eqnarray}
\end{widetext}


\section{A remedy for the infrared divergence}\label{app:remedy-infrared-div}

In our analysis, we had encountered an infrared divergence when calculating 
the Wightman function at a finite temperature.
The occurrence of infrared divergences in Green's functions in spacetime 
dimensions less than $(3+1)$ is not uncommon.
We can turn to the calculation of the Green's functions in $(1+1)$-spacetime 
dimensions to identify possible remedies to regulate the divergence (in this 
context, see, for example, Ref.~\cite{Chowdhury:2019set}). 
Note that, in the case of $(2+1)$-spacetime dimensions, we encounter the 
divergence only when calculating the Green's function at a finite temperature.
(We should clarify that such a divergence does not arise in $(3+1)$-spacetime
dimensions.)
Also, the divergence occurs only in the $m=0$ term in the sum in
Eq.~\eqref{eq:Greenfn-2UDD-thermal-i}.
In the case wherein the detectors are switched on forever, the divergence in
the Wightman function does not affect the transition probability rate of the 
detector, as the $m=0$ term does not contribute [cf. 
Eq.~\eqref{eq:response-fn-infinite-thermal}]. 
We should mention here that such a behavior has also been noticed earlier 
in a related work~\cite{Barman:2021kwg}. 
However, when we consider detectors that are switched on for a finite duration,
the $m=0$ term  in Eq.~\eqref{eq:response-fn-finite-thermal} leads to a non-zero 
contribution and we need to formally regulate the divergence.
We need to do so in such a way that we recover the result in the limit of 
$\bar{\beta}\to\infty$ [viz. Eq.~\eqref{eq:response-fn-finite-Minkowski}], 
i.e. when the temperature of the thermal bath vanishes.
Needless to add, we also need to reproduce our earlier result~\eqref{eq:response-fn-infinite-thermal}
at a finite temperature for the case of detectors that remain switched on 
forever (i.e in the limit $\bar{T}\to \infty$).

Let us first single out the term containing the infrared divergence in the 
finite temperature Wightman function~\eqref{eq:Greenfn-2UDD-thermal-i}. 
Using the following identity (cf. Ref.~\cite{gradshteyn2007}, 8.531.1):
\begin{equation}
J_{0}^2(z)+2\sum_{m=1}^{\infty}\,J_{m}^2(z)=1,
\end{equation}
we can express the Wightman function~\eqref{eq:Greenfn-2UDD-thermal-i} as
\begin{eqnarray}\label{eq:Appn-GreenFn-1}
G^{+}_{\beta}(u) = \mathcal{A}_{0}(u) + \mathcal{A}_{1}(u),
\end{eqnarray}
where the quantities $\mathcal{A}_{0}(u)$ and $\mathcal{A}_{1}(u)$ are given by
\begin{subequations}
\begin{eqnarray}\label{eq:Appn-GreenFn-2}
\mathcal{A}_{0}(u)
&=& \int_{0}^{\infty}\frac{\d q}{4\,\pi}\,
\l(\f{\mathrm{e}^{-i\,\gamma\, q\,u}}{1-\mathrm{e}^{-\beta\, q}}
+\f{\mathrm{e}^{i\,\gamma\, q\,u}}{\mathrm{e}^{\beta\,q}-1}\r),\\
\mathcal{A}_{1}(u)
&=& \int_{0}^{\infty}\frac{\d q}{4\,\pi}\, 
\biggl\{\sum_{\substack{m=-\infty\\ m\neq 0}}^{\infty}
J^{2}_{m}(q\,\sigma)\nn\\ 
& &\times\,\l[\f{\mathrm{e}^{-i\,\gamma\, (q-m\,\Omega)\,u}}{1
-\mathrm{e}^{-\beta\, q}}
+\f{\mathrm{e}^{i\,\gamma\, (q-m\,\Omega)\,u}}{\mathrm{e}^{\beta\,q}-1}\r]~\nn\\
& &-\, 2\,\l(\f{\mathrm{e}^{-i\,\gamma\, q\,u}}{1
-\mathrm{e}^{-\beta\, q}}
+\f{\mathrm{e}^{i\,\gamma\, q\,u}}{\mathrm{e}^{\beta\,q}-1}\r)\nn\\
& &\times\,\sum_{m=1}^{\infty}J^{2}_{m}(q\,\sigma)\biggr\}.
\end{eqnarray}
\end{subequations}
Therefore, the transition probability rate of a detector switched on for a 
finite time through the Gaussian window function can be expressed as
\begin{eqnarray}\label{eq:Appn-RespFn-1}
R_T(\cE) &=& R_T^{0}(\cE)+R_T^{1}(\cE)~,
\end{eqnarray}
where, evidently, $R_T^{0,1}(\cE)$ are given by
\begin{equation}\label{eq:Appn-RespFn-2}
R_T^{0,1}(\cE) = \int_{-\infty}^{\infty}\d u\, \mathrm{e}^{-i\,\cE\,u}\, 
\mathcal{A}_{0,1}(u)\, \mathrm{e}^{-u^2/(2\,T^2)}.
\end{equation}

Since the term~$\mathcal{A}_{1}(u)$ in the finite temperature Wightman function
does not contain the infrared divergence, the corresponding transition probability 
rate $R_T^{1}(\cE)$ can be evaluated as we have done earlier in the other cases.
Therefore, let us turn to the calculation of the rate~$R_T^{0}(\cE)$ that depends 
on the term~$\mathcal{A}_{0}(u)$. 
To do so, let us first explicitly evaluate~$\mathcal{A}_{0}(u)$.
As is often done in the case of $(1+1)$-spacetime dimensions, in order to avoid 
the divergence, we shall take the derivative of~$\mathcal{A}_{0}(u)$ with respect 
to the variable~$u$, thus rendering it safe from the infrared divergence. 
We can then evaluate the integral over $q$ as usual, by introducing an ultraviolet 
regulator of the form of $\mathrm{e}^{-\epsilon\, q}$ to obtain that
\begin{eqnarray}\label{eq:Appn-GreenFn-3}
\f{\partial\mathcal{A}_{0}(u)}{\partial u} 
&=& -\f{i\,\gamma}{4\,\pi\,\beta^2}\, 
\biggl[\psi^{(1)}\l(\frac{i\,\gamma\,u+\epsilon}{\beta}\r)\nn\\
& &-\,\psi^{(1)}\l(\f{-i\,\gamma\,u+\beta+\epsilon}{\beta}\r)\biggr],
\end{eqnarray}
where~$\psi^{(n)}(z)$ denotes the polygamma function of order~$n$. 
Upon integrating over~$u$, we arrive at the expression 
\begin{eqnarray}\label{eq:Appn-GreenFn-4}
\mathcal{A}_{0}(u) &=& -\frac{1}{4\pi\beta}\, 
\biggl[\psi^{(0)}\l(\f{i\,\gamma\,u+\epsilon}{\beta }\r)\nn\\
& &+\, \psi^{(0)}\l(\f{-i\,\gamma\,u+\beta+\epsilon}{\beta}\r)\biggr].
\end{eqnarray}
In the limit of zero temperature (i.e. as $\beta\to \infty$), this expression 
reduces to 
\begin{eqnarray}\label{eq:Appn-GreenFn-5}
\lim_{\beta\to \infty} \mathcal{A}_{0}(u) 
&=& -\f{i}{4\,\pi}\, \f{1}{\gamma\,u-i\,\epsilon}
\end{eqnarray}
which is the result we would have obtained had we taken the limit $\beta\to\infty$ 
in Eq.~\eqref{eq:Appn-GreenFn-2} (i.e. before taking the derivative and carrying
out the integration with respect to variable $u$). 

We can now evaluate the transition probability rate $R_T^{0}(\cE)$ 
of the detector using the expression~\eqref{eq:Appn-GreenFn-4} for 
$\mathcal{A}_{0}(u)$. 
To do so, let us define the dimensionless variable $\tilde{u}=
\gamma\,\Omega\,u$ and parameter $\Bar{\epsilon}= \epsilon\,\Omega$.
Let us also make use of the Fourier transform
\begin{equation}
\mathrm{e}^{-\tilde{u}^2/(2\,\Bar{T}^2)} 
= \f{\Bar{T}}{\sqrt{2\,\pi}}\, \int_{-\infty}^{\infty}
\d\xi\,\mathrm{e}^{i\,\tilde{u}\,\xi-\xi^2\,\Bar{T}^2/2} 
\end{equation}
and the following series expansion of the polygamma function:
\begin{equation}
\psi^{(0)}(z) = -\sum_{k=0}^{\infty}\f{1}{(z+k)}.
\end{equation}
From this last expression, it would be clear that the first and 
the second polygamma functions in Eq.~\eqref{eq:Appn-GreenFn-4} 
would have poles at $\tilde{u}=i\,(k\,\Bar{\beta} +\Bar{\epsilon}) 
=i\,p_{1}(k)$ (i.e. in the upper half of the complex 
$\tilde{u}$-plane) 
and at $\tilde{u}=-i\, [(k+1)\,\Bar{\beta}+\Bar{\epsilon}]=-i\,p_{2}(k)$
(i.e. in the lower half of the complex $\tilde{u}$-plane), respectively. 
The transition probability rate $R_T^{0}(\cE)$ can be written as
\begin{widetext}
\begin{eqnarray}\label{eq:Appn-RespFn-3}
R_T^{0}(\cE) 
&=& \f{\Bar{T}}{4\,\pi\,i\,\,\gamma\,\sqrt{2\,\pi}}\,
\sum_{k=0}^{\infty}\int_{-\infty}^{\infty} 
\d\xi\,\mathrm{e}^{-\xi^2\,\Bar{T}^2/2}
\int_{-\infty}^{\infty}\d \tilde{u}\, 
\mathrm{e}^{i\,(\xi-\cEb)\,\tilde{u}}\,
\l[\f{1}{\tilde{u}-i\,p_{1}(k)}
-\f{1}{\tilde{u}+i\,p_{2}(k)}\r].
\end{eqnarray}
After carrying out the integration over $\tilde{u}$ and imposing 
the appropriate conditions for non-vanishing residues (such as when
$\xi>\cEb$ and $\xi<\cEb$ for the first and the second terms in
the square brackets), we obtain that
\begin{equation}\label{eq:Appn-RespFn-4}
R_T^{0}(\cE) 
= \f{\Bar{T}}{2\,\gamma\sqrt{2\,\pi}}\, \sum_{k=0}^{\infty} 
\biggl[\int_{\cEb}^{\infty}~ \d\xi\,\mathrm{e}^{-\xi^2\,\Bar{T}^2/2}\, 
\mathrm{e}^{-(\xi-\cEb)\,p_{1}(k)}
+\int_{-\infty}^{\cEb} \d\xi\,
\mathrm{e}^{-\xi^2\,\Bar{T}^2/2}\,
\mathrm{e}^{(\xi-\cEb)\,p_{2}(k)}\biggr].
\end{equation}
On further calculating the integral over $\xi$ and then taking 
the limit $\Bar{\epsilon}\to 0$, we arrive at the expression
\begin{eqnarray}\label{eq:Appn-RespFn-4p}
R_T^{0}(\cE) &=&  \f{1}{4\, \gamma} 
\sum_{k=0}^{\infty} \biggl\{\mathrm{e}^{\Bar{\beta}\,k\, 
\l(2\, \cEb\,\bar{T}^2+\bar{\beta}\, k\r)/(2\, \bar{T}^2)}\,
\cE\text{rfc}\l[(\bar{\beta}\, k+\cEb\, \Bar{T}^2)/(\sqrt{2}\, \bar{T})\r]\nn\\
& &+\, \mathrm{e}^{\bar{\beta}\, (k+1) \left(\Bar{\beta}+\Bar{\beta}\, k
-2\, \cEb\, \bar{T}^2\right)/(2\, \bar{T}^2)}\, 
\cE\text{rfc}\l[(\Bar{\beta}+\Bar{\beta}\, k
-\cEb\, \Bar{T}^2)/(\sqrt{2}\, \Bar{T})\r]\biggr\}.
\end{eqnarray}
\end{widetext}

We find that, when the temperature of the thermal bath vanishes (i.e. 
when $\bar{\beta}\to \infty$), in Eq.~\eqref{eq:Appn-RespFn-4}, the quantity 
$p_{2}(k)\to \infty$ for all accessible values of $k$. 
In the same limit, we have $p_{1}(k)=\bar{\epsilon}$ for $k=0$, whereas, 
for the all other values of $k$, we have $p_{1}(k)\to \infty$. 
Therefore, when $\beta\to \infty$, in Eq.~\eqref{eq:Appn-RespFn-4}, 
we are left with only one term of the sum, and it can be expressed as
\begin{eqnarray}\label{eq:Appn-RespFn-5}
\lim_{\substack{\beta\to \infty\\ \Bar{\epsilon}\to 0}} 
R_T^{0}(\cE) 
&=& \f{\Bar{T}}{2\,\gamma\,\sqrt{2\,\pi}}\, 
\int_{\cEb}^{\infty} \d\xi\,\mathrm{e}^{-\xi^2\,\bar{T}^2/2}\,\nn\\
&=& \f{1}{4\,\gamma}\, \mathcal{E}\text{rfc}\l(\f{\cEb\,\bar{T}}{\sqrt{2}}\r).
\end{eqnarray}
In a similar manner, in the $\bar{\beta}\to \infty$ limit, the 
expression~\eqref{eq:Appn-RespFn-4p} is non-zero only when $k=0$.
Also, in this limit, the result reduces to same expression 
as in Eq.~\eqref{eq:Appn-RespFn-5}. 

Furthermore, we can take the limit of infinite interaction time,
i.e. $\Bar{T}\to \infty$) in Eq.~\eqref{eq:Appn-RespFn-4p} and 
observe that the quantity $R_T^{0}(\cE)$ reduces to
\begin{eqnarray}\label{eq:Appn-RespFn-4pp}
R_T^{0}(\cE) &=&  \f{1}{2\, \gamma} \sum_{k=0}^{\infty} 
\mathrm{e}^{-(k+1)\bar{\beta} \,\cEb} 
=  \f{1}{2\, \gamma} \f{1}{\mathrm{e}^{\bar{\beta}\,\cEb }-1}.\quad
\end{eqnarray} 
The same quantity can also be obtained from the second sum of
Eq.~\eqref{eq:response-fn-infinite-thermal} with $m=0$ and by 
utilizing the relation $J_{0}^2(z)=1-2\sum_{m=1}^{\infty}\,J_{m}^2(z)$. 
We also observe that in the zero temperature limit and for 
infinite interaction time, i.e. when $\Bar{T}\to \infty$, 
the quantity $R_T^{0}(\cE)$ vanishes, which is evident 
from Eq.~\eqref{eq:Appn-RespFn-5}. 
(For a different approach to handle this infrared divergence, 
we would refer the reader to Ref.~\cite{Bunney:2023vyj}.)

Let us now provide a similar regularization procedure for the 
thermal Green's function in Eqs.~\eqref{eq:Greenfn-2UDD-thermal}
and~\eqref{eq:W-fn-tb-ub-vb} that, in general, connect two 
detector events. 
In the same manner as in Eq.~\eqref{eq:Appn-GreenFn-1}, we can 
express the Green's function~\eqref{eq:W-fn-tb-ub-vb} as
\begin{eqnarray}\label{eq:Appn-GreenFn-p1}
G^{+}_{\beta_{jl}}(\bar{u},\bar{v}) =
\Bar{\mathcal{A}}_{0{jl}}(\bar{u},\bar{v}) 
+ \Bar{\mathcal{A}}_{1{jl}}(\bar{u},\bar{v}).
\end{eqnarray}
As earlier, in the quantity~$\Bar{\mathcal{A}_{0}}(\bar{u},\bar{v})$, 
we have singled out the contribution containing the infrared divergence.
The quantity $\Bar{\mathcal{A}}_{1}(\bar{u},\bar{v})$ contains all the 
other contributions and it does not diverge in the infrared limit. 
These quantities are given by
\begin{widetext}
\begin{eqnarray}\label{eq:Appn-GreenFn-p2}
\Bar{\mathcal{A}}_{0{jl}}(\bar{u},\bar{v})
&=& \int_{0}^{\infty}\frac{\d q}{4\,\pi}\,
\l[\f{\mathrm{e}^{-i\, \l[\Bar{\alpha}_{1}(q)\,\bar{v} 
+ \Bar{\alpha}_{2}(q)\,\bar{u}\r]/2}}{1-\mathrm{e}^{-\beta\, q}}
+\f{\mathrm{e}^{i\,\l[\Bar{\alpha}_{1}(q)\,\bar{v} 
+ \Bar{\alpha}_{2}(q)\,\bar{u}\r]/2}}{\mathrm{e}^{\beta\,q}-1}\r]~,\nn\\
\Bar{\mathcal{A}}_{1{jl}}(\bar{u},\bar{v}) &=& \int_{0}^{\infty}\frac{\d q}{4\,\pi}\, 
\Bigg\{\sum_{\substack{m=-\infty\\ m\neq 0}}^{\infty}
J_{m}(q\,\sigma_{j})\, J_{m}(q\,\sigma_{l}) \,\l[\f{\mathrm{e}^{-i\, \l[\alpha_{1}(q)\,\bar{v} 
+ \alpha_{2}(q)\,\bar{u}\r]/2}}{1-\mathrm{e}^{-\beta\, q}}
+\f{\mathrm{e}^{i\,\l[\alpha_{1}(q)\,\bar{v} 
+ \alpha_{2}(q)\,\bar{u}\r]/2}}{\mathrm{e}^{\beta\,q}-1}\r]~\nn\\
& &-\, \l[\f{\mathrm{e}^{-i\, \l[\Bar{\alpha}_{1}(q)\,\bar{v} 
+ \Bar{\alpha}_{2}(q)\,\bar{u}\r]/2}}{1-\mathrm{e}^{-\beta\, q}}
+\f{\mathrm{e}^{i\,\l[\Bar{\alpha}_{1}(q)\,\bar{v} 
+ \Bar{\alpha}_{2}(q)\,\bar{u}\r]/2}}{\mathrm{e}^{\beta\,q}-1}\r] 
\l[1-J_{0}(q\,\sigma_{j})\, J_{0}(q\,\sigma_{l})\r]\Bigg\},
\end{eqnarray}
\end{widetext}
where $\Bar{\alpha}_{1}(q)=q\,(\gamma_{j}-\gamma_{l})$ and $\Bar{\alpha}_{2}(q)
=q\,(\gamma_{j}+\gamma_{l})$, i.e. they correspond to $\alpha_{1}(q)$
and $\alpha_{2}(q)$ when $m=0$. 
We can define the corresponding transition probability rates as 
\begin{eqnarray}\label{eq:Appn-RespFn-6}
R_{jl}^T(\cE) &=& R^T_{0{jl}}(\cE)+R^T_{1{jl}}(\cE)~,
\end{eqnarray}
where $R^T_{0{jl}}(\cE)$ depends exclusively on $\Bar{\mathcal{A}}_{0{jl}}(\bar{u},
\bar{v})$ and $R^T_{1{jl}}(\cE)$ on $\Bar{\mathcal{A}}_{1_{jl}}(\bar{u},\bar{v})$. 
We can evaluate the contribution $R^T_{1{jl}}(\cE)$ using the same procedure we 
had adopted in Sec.~\ref{subsubsec:radiative-TB}.
Therefore, we shall now focus only the evaluation of $R^T_{0{jl}}(\cE)$. 
In particular, one can consider general $\gamma_{j}$ and $\gamma_{l}$ for the 
evaluation of $\Bar{\mathcal{A}}_{0{jl}}(\bar{u},\bar{v})$. 
However, recall that, we had considered the same velocities for the two different detectors, 
with different radial distances and angular velocities, to estimate total transition probabilities
(see Figs. \ref{fig:R12-FiniteT-Th-VEb1} and \ref{fig:Ysae-vs-DE-FiniteT-Minkowski}). 
Therefore, for simplicity, we shall set $\gamma_{j}= \gamma_{l}= \gamma$, so that 
$\Bar{\mathcal{A}}_{0{jl}}(\bar{u},\bar{v}) = \mathcal{A}_{0}(\bar{u})$ [cf. Eq.~\eqref{eq:Appn-GreenFn-2}].
In such a case, the expression $R^T_{0{jl}}(\cE)$ will be given exactly by 
Eq.~\eqref{eq:Appn-RespFn-4p} for all $j$ and $l$. 
We can use this result to plot the different transition probability rates as in 
Figs.~\ref{fig:R12-FiniteT-Th-VEb1} and~\ref{fig:Ysae-vs-DE-FiniteT-Minkowski}. 

\bibliographystyle{apsrev}
\bibliography{bibtexfile}

\end{document}